\documentclass[sigconf]{acmart}
\AtBeginDocument{%
  }

\copyrightyear{2025}
\acmYear{2025}
\setcopyright{cc}
\setcctype{by}
\acmConference[SIGIR-AP 2025]{Proceedings of the 2025 Annual International ACM SIGIR Conference on Research and Development in Information Retrieval in the Asia Pacific Region}{December 7--10, 2025}{Xi'an, China}
\acmBooktitle{Proceedings of the 2025 Annual International ACM SIGIR Conference on Research and Development in Information Retrieval in the Asia Pacific Region (SIGIR-AP 2025), December 7--10, 2025, Xi'an, China}\acmDOI{10.1145/3767695.3769502}
\acmISBN{979-8-4007-2218-9/2025/12}
\acmDOI{10.1145/3767695.3769502}

\graphicspath{graphics/}
\usepackage{booktabs}
\usepackage{stfloats}
\usepackage{multirow}
\usepackage[inline]{enumitem}

\newcommand{\header}[1]{\vspace{2mm}\noindent\textbf{#1}}

\newlist{inlinelist}{enumerate*}{1}
\setlist*[inlinelist,1]{label=\roman*),itemjoin={{, }},itemjoin*={{, and }}}

\newcommand{\negskip}{\vspace*{-1.5mm}}

\begin{document}

\title[Investigating LLM Variability in Personalized Conversational Information Retrieval]{Investigating LLM Variability in Personalized \\ Conversational Information Retrieval}


\author{Simon Lupart}
\orcid{0009-0008-2383-4557}
\authornotemark[1]
\affiliation{%
  \institution{University of Amsterdam}
  \city{Amsterdam}
  \country{The Netherlands}}
\email{s.c.lupart@uva.nl}

\author{Daniël van Dijk}
\orcid{0009-0004-2571-842X}
\authornote{Equal contribution.}
\affiliation{%
  \institution{University of Amsterdam}
  \city{Amsterdam}
  \country{The Netherlands}}
\email{daniel.van.dijk@student.uva.nl}

\author{Eric Langezaal}
\orcid{0009-0008-5614-1749}
\authornotemark[1]
\affiliation{%
  \institution{University of Amsterdam}
  \city{Amsterdam}
  \country{The Netherlands}}
\email{eric.langezaal@student.uva.nl}

\author{Ian van Dort}
\orcid{0009-0008-5881-760X}
\authornotemark[1]
\affiliation{%
  \institution{University of Amsterdam}
  \city{Amsterdam}
  \country{The Netherlands}}
\email{ian.van.dort@student.uva.nl}

\author{Mohammad Aliannejadi}
\orcid{0000-0002-9447-4172}
\affiliation{%
  \institution{University of Amsterdam}
  \city{Amsterdam}
  \country{The Netherlands}}
\email{m.aliannejadi@uva.nl}

\begin{abstract}
Personalized Conversational Information Retrieval (CIR) has seen rapid progress in recent years, driven by the development of Large Language Models (LLMs). Personalized CIR aims to enhance document retrieval by leveraging user-specific information, such as preferences, knowledge, or constraints, to tailor responses to individual needs. A key resource developed for this task is the TREC iKAT 2023 dataset, designed to evaluate the integration of personalization into CIR pipelines. Building on this resource, \citet{mo2024pktb_cir} explored several strategies for incorporating Personal Textual Knowledge Bases (PTKB) into LLM-based query reformulation. Their findings suggested that personalization from PTKB could be detrimental and that human annotations were often noisy. However, these conclusions were based on single-run experiments using the commercial GPT-3.5 Turbo model, raising concerns about output variability and repeatability. In this reproducibility study, we rigorously reproduce and extend their work, with a focus on LLM output variability and model generalization. We apply the original methods to the newly released TREC iKAT 2024 dataset, and evaluate a diverse range of models, including Llama (1B to 70B), Qwen-7B, and closed-source models like GPT-3.5 and GPT-4o-mini. Our results show that human-selected PTKBs consistently enhance retrieval performance, while LLM-based selection methods do not reliably outperform manual choices. We further compare variance across datasets and observe substantially higher variability on iKAT than on CAsT, highlighting the challenges of evaluating personalized CIR. Notably, recall-oriented metrics exhibit lower variance than precision-oriented ones, a critical insight for first-stage retrievers, not addressed in the original study. Finally, we underscore the need for multi-run evaluations and variance reporting when assessing LLM-based CIR systems, especially in dense and sparse retrieval or in-context learning settings. By broadening the scope of evaluation across models, datasets, and metrics, our study contributes to more robust and generalizable practices for personalized CIR.
\end{abstract}

\ccsdesc[500]{Information systems~Information retrieval}
\ccsdesc[500]{Information systems~Personalization}
\ccsdesc[500]{Information systems~Language models}
\ccsdesc[500]{Information systems~Users and interactive retrieval}

\keywords{Conversational Information Retrieval; Personalization; Interactive Retrieval; Large Language Model; Conversational Search}

\maketitle

\section{Introduction}

\begin{figure}[t!]
    \centering
    \includegraphics[width=\linewidth]{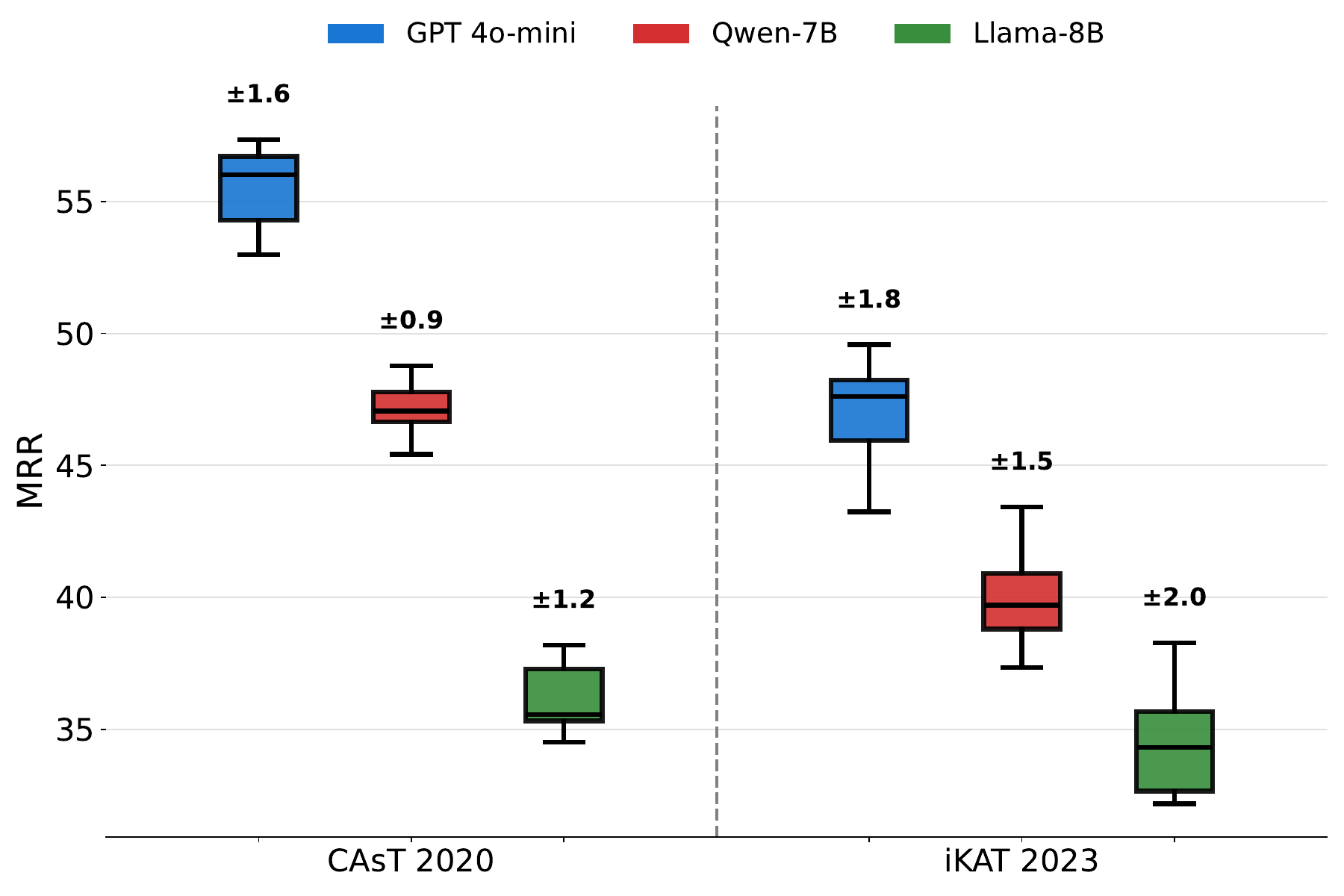}
    \caption{Performance comparison between CAsT 2020 and iKAT 2023, using rewrites from three LLMs. Personalized CIR is more complex, involving not only disambiguations of anaphora and ellipsis but also personalizing from the PTKB. Results averaged over 10 runs, using human PTKB for iKAT.}
    \label{fig:llm_variability}
\end{figure}

Conversational Information Retrieval (CIR) enables users to fulfill their information needs through multi-turn interactions with a retrieval system \cite{hambarde2023information,gao2020recent}. Unlike traditional ad-hoc retrieval, CIR requires models to interpret queries dynamically, considering the evolving context of the conversation. This poses challenges in query understanding, as seemingly similar queries may carry different intents depending on prior dialogue. CIR systems should thus be able to incorporate relevant contextual information from conversation histories while filtering out irrelevant or inaccurate details. This dynamic nature also raises questions about \textit{personalization}, as CIR systems may need to account for user-specific information beyond the immediate conversation to improve retrieval effectiveness.

The TREC iKAT 2023 dataset \cite{trecikat2023} was the first benchmark to introduce personalization in CIR through Personal Textual Knowledge Bases (PTKBs), which encode user-specific attributes in the form of free-text. Each PTKB consists of static textual user details (e.g., dietary preferences, professional background) that persist across conversations and can influence retrieval; they resemble the system's long-term memory about the user. The dataset was designed to foster research and advance the role of personalization in CIR. At the same time, personalization is increasingly integrated into deployed conversational agents (e.g., ChatGPT's memory\footnote{\url{https://openai.com/index/memory-and-new-controls-for-chatgpt}}), reinforcing the importance of continued research in this domain.

A recent study by \citet{mo2024pktb_cir} explored the integration of PTKB statements in CIR by evaluating LLM-based selection methods using the TREC iKAT 2023 dataset. One of the compared approaches relied on query rewriting, where the LLM (GPT-3.5 Turbo) was prompted to reformulate queries to incorporate relevant PTKB information. However, their findings suggest that human selection of relevant PTKB statements (serving as the ground truth) may not correlate with retrieval effectiveness, and excluding PTKB information entirely resulted in better performance. Furthermore, they found that LLMs require explicit guidance to effectively integrate PTKB information, as naive integration can instead degrade retrieval effectiveness. Both those findings challenge common assumptions in personalized CIR, and raise important questions about how user-specific knowledge should be leveraged in current retrieval models.

More broadly, variability in LLMs has often been studied in the field of Natural Language Processing (NLP), where research has shown that LLM outputs can differ significantly even under controlled conditions such as fixed random seeds and zero-temperature decoding \cite{ouyang2025empirical,LLM_evaluation}. This stochastic behavior raises concerns about reproducibility, as even minor variations in a generated text can impact downstream applications. In CIR, LLMs are increasingly used for query rewriting, incorporating relevant information from prior conversation turns and, when applicable, the PTKB. While this technique is crucial for improving retrieval effectiveness, it also introduces LLM variability into CIR pipelines, as slight differences in rewrites can lead to substantial changes in document rankings. Given the growing reliance on LLM-based query reformulation in CIR, bridging the gap between findings on LLM variability in NLP and its implications for retrieval systems ensures that conclusions in personalized CIR research remain robust and reproducible.

As an illustration of this variability, Figure~\ref{fig:llm_variability} presents the performance gap and standard deviation observed on CAsT 2020, a standard conversational search dataset, and iKAT 2023, which incorporates personalization. This margin is due to the complexity of the task. A typical example from CAsT involves simple anaphora resolution: the user query ``How did it govern?'' is reformulated as ``How did the Ottoman Empire govern?'', resolving the pronoun ``it'' to its antecedent (Topic 88, Ottoman Empire). In contrast, in iKAT 2023, the user query ``Which one do you suggest for my mom?'' is reformulated into ``What Turkish souvenir do you suggest for my mom, considering she has a collection of antique crystals and porcelains?'', integrating personal information from the user’s PTKB (Topic 11-2, Finding a Souvenir). The difference in reformulations highlights how the task evolved, with more nuance and reasoning in the case of the personalized rewrite on iKAT. Even minor fluctuations in LLM rewrites can lead to measurable impacts on retrieval performance, especially in personalized settings.

Given these concerns, we systematically reproduce and extend the findings of \citet{mo2024pktb_cir}, under controlled conditions. Specifically, we investigate:
\begin{enumerate}[label=\textbf{RQ\arabic*}]
    \item \label{int:rq1} How does LLM variability impact the findings of \citet{mo2024pktb_cir} on the integration and usefulness of PTKB in CIR?
    \item \label{int:rq2} Do the findings of \citet{mo2024pktb_cir} hold consistently for open-source LLMs of varying sizes?
    \item \label{int:rq3} To what extent do these findings generalize to the new TREC iKAT 2024 personalized CIR dataset?
\end{enumerate}

To ensure a rigorous and reproducible evaluation, we introduce the following methodological improvements:
\begin{itemize}[leftmargin=*]
    \item \textbf{Mitigating LLM variability} by averaging results across multiple runs to obtain a more reliable measure of retrieval effectiveness \citep{ouyang2025empirical}.
    \item \textbf{Comparing open-source LLMs} (Llama, Qwen) of varying sizes, with closed-source GPT, ensuring deterministic query rewrites under fixed seeds.
    \item \textbf{Extending evaluation to TREC iKAT 2024}, testing if prior conclusions generalize across datasets.
    \item \textbf{Including Recall@1000 as an additional metric}, improving the evaluation of first-stage retrieval models.
\end{itemize}

Our findings show that the original conclusions from \citet{mo2024pktb_cir} are not fully reproducible, emphasizing the need to account for LLM variability in CIR evaluations. Instead, we show that the PTKB statements that humans judged to be relevant generally align with ranking performance, improving retrieval in both TREC iKAT 2023 and 2024. While the smaller open-source LLMs usually underperform compared to commercial GPT models, we observe that when guided with in-context learning examples, they perform within a reasonable range, compared to GPT-based models. On the other hand, larger open-source LLMs like Llama-3.3-70B perform on par with some closed-source ones. Despite all this, all LLMs seem to be subject to variability. We also observe that the performance variability is generally lower on the newly added recall-oriented metrics, an important finding given that all compared methods from both the original work and our reproduction are first-stage retrieval models. We contribute by releasing our complete Python codebase on Github \url{https://github.com/EricLangezaal/PersonalizedCIR}.

\section{Related Work}

\header{Query Rewriting in Conversational IR.} Conversational Information Retrieval (CIR) opposes itself to ad-hoc search from the conversational nature of the input. While queries in traditional IR are usually limited to a few words, user interactions in CIR systems evolve into multi-turn conversations~\cite{treccast19,frameworkcs,neuralcs}. Modeling the context and information needs of the user is therefore the main challenge associated with CIR, together with creating denoised representations~\cite{convdr,lupart2025disco}. In this context, query rewriting aims at generating a standalone self-contained query based on the original conversation history and last user utterance~\cite{ye-etal-2023-enhancing,elgohary-etal-2019-unpack}. This method has developed significantly with the growth of decoder models and later with LLMs in zero-shot and few-shot fashion~\cite{jin2023instructor, mao2024chatretriever, mao-etal-2023-large, mao2022curriculum, qu2020open}. To further enhance rewrite quality and semantic coverage, several recent studies have explored prompting LLMs to generate multiple rewrites per turn~\cite{mao-etal-2023-large,abbasiantaeb2024generatingmultiaspectqueriesconversational,lupart2024irlab,mo-etal-2024-chiq}. In particular, LLM4CS~\cite{mao-etal-2023-large}, leverages the stochasticity of LLMs by prompting multiple times and aggregating diverse rewrites, reporting performance improvements through greater lexical and semantic variation. However, despite these gains, prior work has not systematically examined the nature or variability of the generated rewrites themselves. Moreover, since all these methods rely on LLMs as generative backbones, they inherently inherit the variance and instability associated with LLM outputs. Our study addresses this gap by providing a detailed analysis of LLM rewrite variability, establishing its impact on retrieval performance and its implications for robust CIR evaluation.

\header{Personalized CIR.} 
In other textual domains, recent studies have proposed methods for integrating~\cite{salemi2023lamp} and evaluating~\cite{salemi2025expert, kumar2024longlamp} personalized information in LLM-based query reformulation. In contrast, most prior work in conversational IR does not incorporate external personalized information into the reformulation process, relying solely on the conversational history to infer user intent. Yet in real-world applications, additional user-specific information (e.g. preferences, goals, or constraints) is often available and could be leveraged to improve retrieval quality. This raises the question of how to effectively incorporate such information into query reformulation. Building on TREC iKAT, \citet{mo2024pktb_cir} is among the first studies to compare the effect of personalization in CIR, by passing (a selection of) user information to the LLM reformulation task, hence our motivation to reproduce their work. In relation to the query rewriting task, complementary evidence is provided by MQ4CS~\cite{abbasiantaeb2024generatingmultiaspectqueriesconversational}, which reports that TREC iKAT 2023 benefits more from generating multiple rewrites than TREC CAsT. They attribute this to the greater complexity of information needs in personalized CIR settings. This further underscores the importance of revisiting and reproducing the findings of \citet{mo2024pktb_cir}, to better understand how personalization and rewrite variability interact in such contexts.

\header{LLM Variability.}
While LLMs allow for powerful reformulation, a major challenge for reliable evaluation is that they commonly introduce variance in results due to their stochastic nature.
This is especially troublesome for commercial LLMs such as OpenAI's GPT models, where the API-based access limits both the available information and control over inference conditions. Previous research has demonstrated major variance across LLM outputs \cite{LLM_evaluation, ouyang2025empirical}, even when provided with identical inputs \cite{atil2024llmstabilitydetailedanalysis}. 
LLMs often include a temperature parameter, which could influence the degree of variability when sampling outputs. While a lower temperature value could lower the variability, this could also somewhat degrade performance \cite{temperature}. However, specifically for GPT models, it has been demonstrated that even with a temperature of zero and a fixed seed, these models are not fully deterministic~\cite{chann2023, ouyang2025empirical}. While non-determinism can be beneficial for many tasks, it complicates the evaluation of commercial models. In contrast, open-source models, which can be run locally, can be guaranteed to be deterministic under a fixed seed.

Beyond general observations of stochasticity, recent studies have also investigated LLM behavior under more complex and challenging conditions. \citet{qi2024quantifying} reported increased variance on out-of-domain tasks compared to in-domain ones, across models of varying sizes. \citet{cao2024worst} similarly found that LLM performance fluctuates substantially with prompt phrasing, and that increasing model size does not necessarily mitigate this variability. Finally, \citet{baidya2025behavior} showed that LLM agents exhibit behaviors that diverge significantly from human reasoning, with discrepancies increasing as task complexity rises or model size decreases. Despite these findings and the growing reliance on LLMs in CIR, no prior work has examined how such variability manifests in the more complex setting of Personalized CIR. Our study fills this gap by systematically analyzing LLM-induced variability in personalized query rewriting and its impact on retrieval.

\section{Methodology}
\label{sec:methodology}

This section aims to highlight the different techniques developed in the original paper, to summarize the overall pipeline, and to illustrate which parts have been extended for this reproduction study. Borrowing notation from \cite{mo2024pktb_cir}, the goal in conversational information retrieval is to retrieve relevant passages $d^+ \in D$ based on user $u$’s current conversation turn $q_n^u$ and the conversation history $\mathbb{H} = {q_i}_{I=1}^{n-1}$. Specifically for personalized CIR, this problem definition is extended by also making the search dependent on the personal textual knowledge base (PTKB) \cite{trecikat2023} $\mathbb{U}={s_t}_{t=1}^{T}$, with $s_t$ being the t-th sentence in the PTKB. 

The broad approach is to first select the relevant part of the PTKB, using various approaches and baselines. Next, the current query turn is reformulated into a standalone query by providing the conversation history and selected PTKB to a LLM. This reformulated self-contained query is then used in either sparse or dense passage retrieval, after which the retrieval performance can be calculated.

\begin{table}[t!]
    \centering
    \caption{Statistics for the personalized iKAT datasets, and the non personalized CAsT dataset.}
    \label{methodology:tab:trec}
    \begin{tabular}{l|c|c|c}
        \toprule
        \multirow{2}{*}{\textbf{Statistic}} & \textbf{CAsT 2020} & \textbf{iKAT 2023} & \textbf{iKAT 2024} \\
         & \textbf{Test} & \textbf{Test} & \textbf{Test} \\
        \midrule
        Topics & 25 & 13 & 14 \\
        Conversations & 25 & 25 & 16 \\
        Assessed Turns & 216 & 176 & 116 \\
        Turns with PTKB & - & 63 & 54 \\
        Passage collection & 38 Million & \multicolumn{2}{c}{116.8 Million} \\
        \bottomrule
    \end{tabular}
\end{table}

\subsection{TREC iKAT Datasets}

To assess the effect of personalized information in the retrieval task, a dedicated dataset is required. The TREC iKAT 2023 dataset \cite{trecikat2023} contains 36 conversations about 21 different topics. Each conversation consists of multiple turns, a subset of which has annotations such that they can be assessed for information retrieval, using a subset of passages originating from the ClueWeb22-B corpus \cite{overwijk2022clueweb22}. For each conversation, the dataset also features roughly ten PTKB elements containing information about the user. Unlike similar CIR datasets, iKAT 2023 contains two types of human-annotated relevance judgments. Firstly, it contains query-passage relevance judgments for each assessed turn, which can be used to assess the search performance. Secondly, for each assessed turn, it lists all PTKB sentences that the annotators deemed relevant for that turn/query, if any. For an overview of the statistics of this dataset, please see Table~\ref{methodology:tab:trec}. Experiments for reproduction will initially be conducted on iKAT 2023, but we have also extended this work by including results for the updated iKAT TREC 2024 dataset~\cite{trecikat2024}. 

\subsection{Personalized Query Rewriting}

\begin{figure}[t!]
    \centering
    \includegraphics[width=\linewidth]{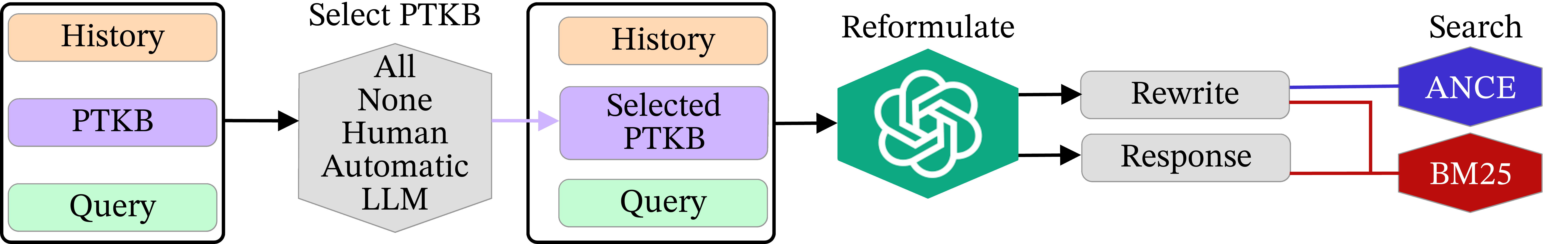}
    \caption{Outline of the LLM-aided conversational IR pipeline with a separate PTKB selection and reformulation stage. Select Then Retrieve (STR) follows this setup.}
    \label{fig:methodology:str}
    \Description{Outline of the LLM-aided conversational IR pipeline with a separate PTKB selection and reformulation stage.}
\end{figure}

\begin{figure}[t!]
    \centering
    \includegraphics[width=0.9\linewidth]{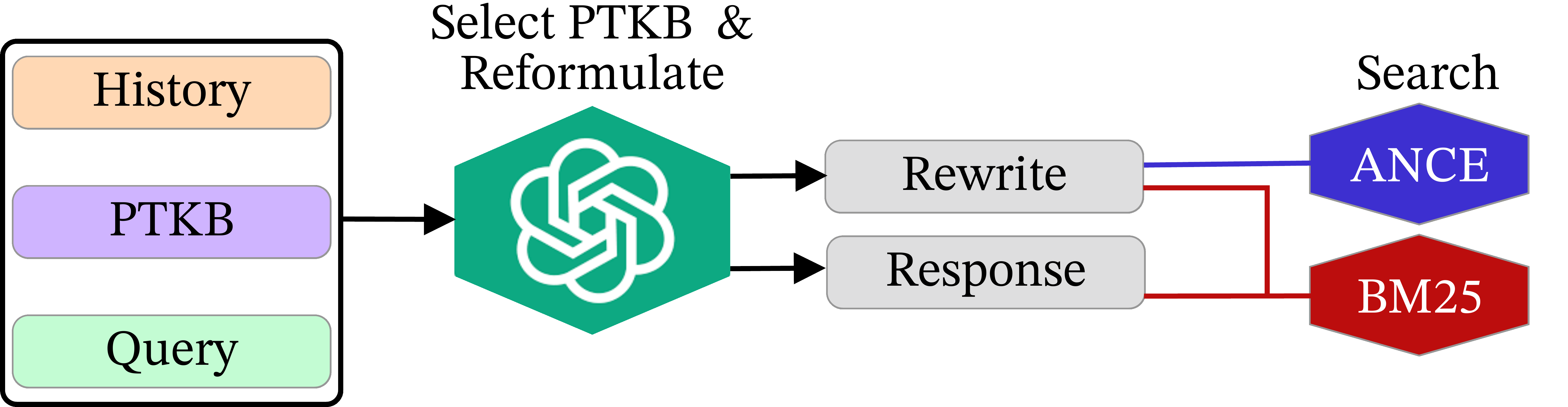}
    \caption{Outline of the Select And Reformulate (SAR) conversational IR pipeline.}
    \label{fig:methodology:sar}
    \Description{Outline the of Select And Reformulate (SAR) conversational IR pipeline.}
\end{figure}

In personalized CIR, the history, current query and a (selection of) the PTKB need to be transformed into a representation suitable for passage retrieval. Firstly, the authors devise multiple methods to select the relevant PTKB for a given conversation turn. They also experiment with two non-selective baselines. After the PTKB selection, the original paper always leverages OpenAI’s GPT-3.5 (gpt-3.5-turbo-16k) to create a single self-contained query through a reformulation prompt. The authors also experiment with in-context learning in the reformulation step, where they use different numbers of examples from the training set as examples for the LLM. The next sections outline the various approaches developed for PTKB processing and query reformulation in more detail. Figures~\ref{fig:methodology:str} and ~\ref{fig:methodology:sar} provide an overview of the pipelines, where the PTKB selection is followed by a separate reformulation stage.

\header{PTKB Selection.}
The authors distinguish between five approaches that separately process the PTKB items:
\begin{itemize}[leftmargin=*]
    \item \textbf{Select Then Reformulate (STR)}: This method first uses an LLM to select relevant PTKB items from the full list of PTKBs, and then reformulates the last user utterance using only the selected subset. This approach decomposes the task into two distinct inference steps (LLM calls), breaking down the task to simpler ones, while being more interpretable.
    \item \textbf{Select and Reformulate (SAR)}: This method uses a single LLM call to jointly perform both PTKB selection and query reformulation. While the approach is more straightforward and efficient than STR, it does not explicitly expose which PTKB statements were used in the rewrite. This implicit reasoning process can become problematic as the size and complexity of the PTKB increase, potentially making it harder for the model to identify and incorporate relevant information effectively.
    \item \textbf{None}: This baseline consists of simply not using any PTKB items for any query, regardless of annotated relevance, resulting in a non-personalized rewrite, except in cases where the user information has already been revealed in the history of the conversation.
    \item \textbf{Use All}: This baseline provides the full set of PTKB statements to the LLM during query reformulation, without performing any prior selection. It is conceptually similar to the SAR method, with the primary difference being a prompt that explicitly states to include all PTKB content in the rewrite rather than prompting the model to infer relevance while rewriting.
    \item \textbf{Human}: The Human method uses PTKB statements marked as relevant for that specific query. However, the authors argue that there might be a discrepancy between PTKB informativeness and actual improvements for retrieval effectiveness. While a certain PTKB item might seem relevant for a query to a human annotator, its conjunction might not actually be related (enough) to any of the passages to influence retrieval.
    \item \textbf{Automatic/Oracle}: Lastly, the Automatic method reformulates a query with each PTKB sentence separately, resulting in approximately ten reformulations per query (corresponding to the average PTKB set size). Among these reformulations, the combination of the query and PTKB element leading to the highest NDCG@3 score is then used for calculating retrieval performance. We renamed this method \textit{Oracle} as it uses the relevance judgement to select the unique best PTKB. A more detailed discussion is provided in Section~\ref{sec:analysis:oracle}.
\end{itemize}

\negskip

\header{LLM Query Rewriting.} \label{sec:meth:prompt}
Once the PTKB has been processed, either by selecting relevant statements, using all or omitting them, it is combined with the history and current conversation turn to generate the LLM-based rewrite. This rewritten query is then used as input to the retrieval model to obtain relevant passages. To further augment the reformulating, the authors prompt the LLM to generate up to five rewrites, which are concatenated to form the final query. This acts as a form of query augmentation, introducing synonyms to improve coverage and retrieval effectiveness. For BM25, the author also generates a preliminary answer (\textit{response}) alongside the rewrite. They argue that leveraging the LLM's internal knowledge may improve consistency and accuracy between the rewritten queries and the generated responses~\cite{mo2024pktb_cir}. We thus followed those settings for our query reformulation experiments to ensure comparability with prior results.

\subsection{LLM Generalizability and Stability}
\label{methodology:reformulation:extension}
\header{Variance.} While reproducing the experiments, we noticed substantial challenges regarding the stability of the results, with metric differences of multiple percentage points across identical runs. Given that prior research has shown that GPT models are inherently stochastic regardless of seeding and temperature settings, and realizing that altering the temperature might impact results, we choose to use the same (default) temperature value as the original authors.
To increase reliability, we opted to extend this work by repeating each experiment several times (five to ten times), such that we could obtain robust estimates of the means and variances of the suggested methods. We reproduced the main results of the original authors first using GPT-3.5 Turbo (gpt-turbo-3.5-16k) to strictly follow their experimental setup, and then opted to continue and extended the additional experiments with the OpenAI GPT-4o Mini (2024-07-18) model, which outperforms GPT-3.5 Turbo while being roughly 20 times cheaper for input tokens\footnote{\url{https://openai.com/index/gpt-4o-mini-advancing-cost-efficient-intelligence/}}\footnote{\url{https://platform.openai.com/docs/models}}. Using this model, we were able to repeat all experiments in the original paper fivefold. 

\header{Open-source LLMs.} In email correspondence, the original authors acknowledged the variability issue and mentioned they attempted to use an open-source LLM with 7B parameters. However, they found that this model failed to match the performance of GPT-3.5. We opted to also investigate this claim with an open-source LLM of similar size. For this goal, we ran the open-source Llama-3.1-8B~\cite{llama3} and Qwen-2.5-7B~\cite{hui2024qwen2} models five times for the SAR methods using the same prompts as GPT, each time with a different seed. Consequently, the average performance and variance across the five runs with different seeds are exactly reproducible. Although all experiments of the original paper could be repeated with Llama, this ablation primarily serves as a proof of concept. Therefore, we tested the model on the methods most influenced by the choice of LLM (SAR method), in zero-shot and few-shot settings. This would also allow for comparing the learning capacities of the LLMs when given few-shot examples.

\header{Parameter Size.} To further investigate and benchmark the current performance of open-source LLMs, we experiment with LLMs of various sizes. We compare models from 1B parameters to up to 70B. For this, we use the latest Llama-3 family of models, with 1B, 3B, 8B, and 70B parameters~\cite{llama3}.

\subsection{Passage retrieval}
In the previous section, a method has been developed to process the conversation history, current turn and (relevant) PTKB through an LLM. This results in a self-contained query (\textit{rewrite}), and a preliminary response also generated by the LLM. The authors experiment with two vastly different passage retrieval strategies:

\begin{itemize}[leftmargin=*]
    \item \textbf{Sparse BM25:} This technique constructs an inverted index of the entire passage collection, mapping each vocabulary word to the passages it occurs in. A lexical search can then be conducted on the search query, performing exact word matching, which means BM25 is not robust to synonyms. It should be noted that by construction, this technique favors lengthy queries with multiple variations of a sentence, as that increases the likelihood of an exact word match. The original paper implicitly benefits from this by opting to use a concatenation of the \textit{rewrite} and the \textit{response} to perform BM25 search. The index and search were implemented in Pyserini~\cite{lin2021pyserini} with $k_1=0.82$ and $b=0.68$. 
    \item \textbf{Dense ANCE:} Next, the authors experiment with vector-based search through the dense ANCE \cite{ANCE} neural encoder. After embedding every passage into a fixed-length vector, the query embedding can be matched with passage embeddings based on cosine similarity, implemented efficiently in Faiss \cite{faiss}. Unlike BM25, this technique does not rely on exact lexical matching, and therefore does not necessarily benefit from longer queries, instead favoring more meaningful or concise queries. The authors attempt to accommodate for this by only using the \textit{rewrite}'s embedding for ANCE. 
\end{itemize}

\subsection{Evaluation}
\label{sec:methodology:evaluation}

To evaluate the effectiveness of a certain reformulation technique and either sparse or dense retrieval, the authors first calculate metric results for each query. This is achieved by comparing the retrieved passages with human-labeled ground truth relevance scores for passages for each assessed query. They compute the binary metrics Mean Average Precision (MAP) and Mean Reciprocal Rank (MRR), which only take into account if a passage has a relevance score of 1 or more. Furthermore, they compute the Normalized Discounted Cumulative Gain (NDCG) at cutoffs of 3 (NDCG@3) and 5 (NDCG@5) passages. These latter two metrics take into account the actual numerical relevance too, which is a number between zero (not relevant) and four (high relevance). Unlike the original paper, we also include recall at a cutoff of 1000 (R@1000) in our evaluation, as this metric is not only common for iKAT~\cite{trecikat2023} but also crucial for assessing the performance of first-stage retrieval methods like BM25 and ANCE~\cite{recall_neural_ir}. The high cutoff provides an interesting point of comparison, particularly given the low cutoff of most other metrics reported. After scoring the performance for each query, the authors simply average the results for each metric across all queries in the dataset.  
When investigating the significance of claims, we utilize a paired T-Test with a p-value of $0.05$ using the means and standard deviations reported.

\section{Experimental Results}
This section will introduce the main experimental results of the original paper and outline our attempted reproduction and subsequent extensions. To answer our proposed research questions, the results are divided accordingly. Sections~\ref{sec:results:repro_main} and \ref{sec:results:icl} attempt to reproduce both the general PTKB selection and specific in-context learning experiments of the original paper, respectively. A brief reflection on the automatic method is also provided. Next, Section~\ref{sec:results:llama} extends the original paper by comparing results with the open-source Llama and Qwen models instead of GPT-3.5. Lastly, Section~\ref{sec:results:ikat2024} answers our third research question, reproducing the findings on the iKAT 2024 dataset, to assess the generalization of the proposed approaches.

\label{sec:results:rq1}
\begin{table*}
  \caption{Results for the baseline methods on the iKAT 2023 full test set (176 turns). The best method per metric is highlighted in bold, a dagger (\textsuperscript{\textdagger}) indicates statistical significance (paired T-test p-value < 0.05) compared to using no PTKB. }
  \label{tab:fullset_comparison}
  \begin{tabular}{lllclllll}
    \toprule
    \textbf{Retrieval} & \textbf{PTKB} & \textbf{LLM} & \textbf{Runs} & \textbf{MRR} & \textbf{NDCG@3} & \textbf{NDCG@5} & \textbf{R@1000} & \textbf{MAP} \\
    \midrule
      \multirow{18}{*}{BM25} & None & GPT 3.5 (theirs) & 1 & 44.35 & 21.22 & 20.68 & - & 8.91 \\
      & None & GPT 3.5 (ours) & 5 & 36.11 \textpm 1.75 & 16.27 \textpm 0.81 & 16.42 \textpm 0.65 & 41.42 \textpm 0.45 & 8.37 \textpm 0.37 \\
      & None & GPT-4o mini (ours) & 5 & 45.75 \textpm 1.14 & 20.95 \textpm 0.96 & 20.46 \textpm 0.60 & 47.09 \textpm 0.70 & 9.72 \textpm 0.27 \\
        \cmidrule{2-9}
      & Use all & GPT 3.5 (theirs) & 1 & 40.36 & 19.19 & 18.84 & - & 8.28 \\
      & Use all & GPT 3.5 (ours) & 5 & 35.94 \textpm 2.19 & 16.74 \textpm 0.92 & 16.42 \textpm 0.48 & 41.32 \textpm 0.33 & 8.04 \textpm 0.22 \\
      & Use all & GPT-4o mini (ours) & 5 & 45.45 \textpm 1.40 & 21.22 \textpm 0.62 & 20.20 \textpm 0.36 & 47.54 \textpm 0.76 & 9.34 \textpm 0.11 \\
      \cmidrule{2-9}
      & Human & GPT 3.5 (theirs) & 1 & 41.65 & 19.66 & 19.46 & - & 8.82 \\
      & Human & GPT 3.5 (ours) & 5 & 38.07 \textpm 1.52 & 17.82 \textpm 1.12 & 17.09 \textpm 0.70 & 40.97 \textpm 0.96 & 8.26 \textpm 0.11 \\
      & Human & GPT-4o mini (ours) & 5 & \textbf{45.94} \textpm 1.97 & \textbf{21.58} \textpm 1.35 & \textbf{20.98} \textpm 1.15 & \textbf{47.99}\textsuperscript{\textdagger} \textpm 0.37 & \textbf{9.80} \textpm 0.31 \\
      \cmidrule{2-9}
      & STR 0-shot & GPT 3.5 (theirs)& 1 & 41.53 & 18.96 & 18.09 & - & 8.37 \\
      & STR 0-shot & GPT 3.5 (ours) & 5 & 39.08 \textpm 2.22 & 17.91 \textpm 1.60 & 17.52 \textpm 1.17 & 40.53 \textpm 0.96 & 8.37 \textpm 0.38 \\
      & STR 0-shot & GPT-4o mini (ours) & 5 & 44.66 \textpm 1.34 & 21.10 \textpm 1.01 & 20.05 \textpm 0.69 & 46.73 \textpm 0.65 & 9.33 \textpm 0.21 \\
      \cmidrule{2-9}
      & SAR 0-shot & GPT 3.5 (theirs) & 1& 36.04 & 17.48 & 16.87 & - & 8.02 \\
      & SAR 0-shot & GPT 3.5 (ours) & 5 & 38.99 \textpm 1.31 & 17.58 \textpm 1.39 & 16.73 \textpm 0.95 & 41.24 \textpm 0.76 & 7.51 \textpm 0.23 \\
      & SAR 0-shot & GPT-4o mini (ours) & 5 & 40.58 \textpm 1.99 & 18.54 \textpm 0.91 & 17.68 \textpm 0.60 & 39.64 \textpm 0.57 & 7.35 \textpm 0.13 \\
      \cmidrule{2-9}
      
      \cmidrule{2-9}
      & Automatic/Oracle & GPT 3.5 (theirs) & 1 & 40.29 & 19.12 & 18.87 & - & 8.58 \\
      & Automatic/Oracle & GPT-4o mini (ours)& 5 & 64.36 \textpm 0.99 & 37.33\textpm 0.35 & 31.27\textpm 0.06 & 43.10 \textpm 0.44 & 10.24\textpm 0.24 \\
    \bottomrule
  \end{tabular}
\end{table*}

\subsection{Reproduction and LLM Variability}
\label{sec:results:repro_main}
To answer \ref{int:rq1} on the reproducibility of the findings of the original authors, we focus in this section on all six PTKB selection methods, evaluated on the TREC iKAT 2023 dataset. We compare both the results of the original paper and the aggregated results of our fivefold experiments using GPT-3.5 Turbo and GPT-4o mini. 

Table~\ref{tab:fullset_comparison} presents the retrieval results for BM25, while Table~\ref{tab:fullset_ance} reports those for ANCE. As explained in our methodology, GPT-reformulated queries vary substantially when the experiments are repeated multiple times, leading to noticeable variance in retrieval performance. When comparing the original GPT 3.5 results with our fivefold repeated runs, we observe some differences that fall within the range of the standard deviation, while others exceed it, highlighting the sensitivity of the pipeline to LLM variability. Despite this variability, the authors state that the \textit{None} baseline (i.e. excluding PTKB information) achieves the highest scores across all metrics, with ``significantly better MRR, NDCG@3, and NDCG@5''. However, our results do not support this conclusion on either GPT-3.5 or GPT-4o mini. Specifically, GPT-4o mini performance with BM25 and Human annotations outperforms all other selection methods (including the \textit{None} and \textit{Use all} baselines) on all metrics. Similarly, for GPT 3.5, the Human selection achieves better performance than the \textit{None} baseline on most metrics. While our results do not reproduce the original findings, this divergence is consistent with expectations, as the iKAT dataset and its human annotations were explicitly constructed to support personalization in CIR. For ANCE, human annotations do not score best for some metrics, but there is no significant difference with either using all or no PTKB. Finally, consistent with prior findings, we observe that sparse BM25 retrieval continues to outperform dense ANCE across all conditions.

\begin{table}
  \caption{ANCE Results for the baseline methods on the iKAT 2023 full test set (176 turns). Our runs are a fivefold average using GPT-4o mini. Statistical significance (paired T-test p-value < 0.05) compared to using no PTKB with (\textsuperscript{\textdagger}).}
  \label{tab:fullset_ance}
  \centering
  \begin{tabular}{lccc}
    \toprule
    \textbf{Method} & \textbf{MRR} & \textbf{NDCG@3} & \textbf{R@1000} \\
    \midrule
    \multicolumn{4}{c}{\textbf{Dense retrieval (ANCE)}} \\
    \midrule
    None (theirs) & 32.47 & 14.25 & - \\
    None (ours) & 38.08 \textpm 2.02 & 18.62 \textpm 1.02 & 40.02 \textpm 0.10 \\
    \midrule
    Use all (theirs) & 33.64 & 15.30 & - \\
    Use all (ours) & \textbf{38.94} \textpm 0.87 & \textbf{18.90} \textpm 1.36 & 40.43 \textpm 0.38 \\
    \midrule
    Human (theirs) & 33.63 & 15.98 & - \\
    Human (ours) & 38.41 \textpm 0.92 & 18.44 \textpm 0.65 & \textbf{41.21}\textsuperscript{\textdagger} \textpm 0.38 \\
    \midrule
     STR 0-shot (theirs) & 32.37 & 15.05 & - \\
      STR 0-shot (ours) & 37.33 \textpm 1.77 & 17.99 \textpm 0.79 & 40.57 \textpm 0.50 \\
      \midrule
      SAR 0-shot (theirs)& 31.76 & 14.78 & -\\
      SAR 0-shot (ours) & 33.50 \textpm 0.83 & 15.67 \textpm 1.01 & 36.31 \textpm 0.27  \\
    \midrule
    \midrule
     Oracle (theirs) & 31.08 & 14.36 & - \\
    Oracle (ours) & 55.37\textpm 1.52 & 31.55 \textpm 1.19 & 39.42 \textpm 0.73 \\
    \bottomrule
  \end{tabular}
\end{table}

Now looking at the added metric, Recall@1000, we find that interestingly, Human PTKB selection on GPT-4o mini significantly outperforms the \textit{None} baselines on both ANCE and BM25. We also observe that Recall@1000 has an overall lower variance than precision-oriented metrics. This aligns with common intuition: precision metrics like NDCG@3 are highly sensitive to the top-ranked results, which can fluctuate considerably depending on small variations in the rewrite. In contrast, recall at a larger cutoff reflects a broader coverage of relevant content and is more stable across repeated runs.  Including this high-cutoff recall metric provides a more robust basis for comparison across methods, especially given that all evaluated systems perform first-stage retrieval. In this context, evaluating recall is arguably more appropriate than relying solely on early-precision metrics~\cite{recall_neural_ir}.

\header{The Automatic/Oracle Method.} The automatic method benefits extensively from GPT’s large output variance, as it selects the highest-NDCG@3-scoring reformulation from roughly ten query reformulations (one for each PTKB element). Consequently, this method is expected to dominate the other methods on NDCG@3, on which it was optimized, and any metric at low cutoffs like MRR or NDCG@5. Table~\ref{tab:fullset_comparison} and Table \ref{tab:fullset_ance} indeed confirm that this method consistently outperforms the others by a major margin on all metrics except for Recall@1000, where it actually underperforms significantly. While this method perform very well on NDCG@3, the poor performance on recall highlights that it is only able to do a post-hoc optimization on the top-3 ranked results. We provide a further analysis of the Automatic method in Section~\ref{sec:analysis:oracle}.

\label{sec:results:rq2}
\begin{table*}[t]
  \caption{Our reproduction of in-context learning results for iKAT 2023 full test set (176 turns). These fivefold averages have been obtained with GPT-4o mini. Bold indicates best metric performance for all STR and SAR methods.}
  \label{tab:icl_fullset_comparison}
  \begin{tabular}{lccccccc}
    \toprule
    \textbf{Model} & \textbf{Method} & \textbf{Sample} & \textbf{MRR} & \textbf{NDCG@3} & \textbf{NDCG@5} & \textbf{R@1000} &\textbf{MAP} \\
    \midrule
      \multirow{10}{*}{BM25}  & None & - & 45.75 \textpm 1.14 & 20.95 \textpm 0.96 & 20.46 \textpm 0.60 & 47.09 \textpm 0.70 & 9.72 \textpm 0.27 \\
       & Human & - & 45.94 \textpm 1.97 & 21.58 \textpm 1.35 & 20.98 \textpm 1.15 & 47.99 \textpm 0.37 & 9.80 \textpm 0.31 \\
      \cmidrule{2-8}
      & \multirow{4}{*}{STR} & 0-shot & 44.66 \textpm 1.34 & 21.10 \textpm 1.01 & 20.05 \textpm 0.69 & 46.73 \textpm 0.65 & 9.33 \textpm 0.21 \\
      & & 1-shot & 44.99 \textpm 1.86& \textbf{21.10} \textpm 1.06&\textbf{20.29} \textpm 0.90 & \textbf{47.25} \textpm 0.70 & \textbf{9.55} \textpm 0.10\\
      & & 3-shot & \textbf{45.15} \textpm 2.73&20.49 \textpm 1.42&19.73 \textpm 1.03& 47.18 \textpm 1.13 &9.26 \textpm 0.28\\
      & & 5-shot & 43.59 \textpm 0.37&19.92 \textpm 0.64&19.45 \textpm 0.66& 46.93 \textpm 0.71 &9.13 \textpm 0.25\\
      \cmidrule{2-8}
      & \multirow{4}{*}{SAR} & 0-shot & 40.58 \textpm 1.99 & 18.54 \textpm 0.91 & 17.68 \textpm 0.60 & 39.64 \textpm 0.57 & 7.35 \textpm 0.13 \\
      & & 1-shot & 40.72 \textpm 2.21 & 18.97 \textpm 1.59&18.12 \textpm 1.05& 41.79 \textpm 0.38 &7.84 \textpm 0.23\\
      & & 3-shot & 40.36 \textpm 1.59&17.70 \textpm 0.74&17.24 \textpm 0.50& 41.70 \textpm 0.81 &7.38 \textpm 0.31\\
      & & 5-shot & 39.82 \textpm 1.20&17.80 \textpm 1.34&17.45 \textpm 0.84 & 42.70 \textpm 0.79 &7.77 \textpm 0.22\\
    \midrule
    \midrule
      \multirow{10}{*}{ANCE}  & None & - &  38.08 \textpm 2.02 & 18.62 \textpm 1.02 & 17.97 \textpm 0.72 & 40.02 \textpm 0.10 & 7.67 \textpm 0.30 \\
      & Human & - & 38.41 \textpm 0.92 & 18.44 \textpm 0.65 & 17.84 \textpm 0.57 & 41.21 \textpm 0.38 & 7.77 \textpm 0.30 \\
      \cmidrule{2-8}
      & \multirow{4}{*}{STR} & 0-shot & 37.33 \textpm 1.77 & 17.99 \textpm 0.79 & 17.02 \textpm 0.70 & \textbf{40.57} \textpm 0.50 & 7.25 \textpm 0.18 \\
      & & 1-shot & 37.04 \textpm 2.52&17.15 \textpm 0.89&16.70 \textpm 0.51& 40.05 \textpm 0.59 &7.26 \textpm 0.12 \\
      & & 3-shot & 36.73 \textpm 1.81&18.04 \textpm 0.96&17.23 \textpm 0.54& 40.14 \textpm  0.56 &7.23 \textpm 0.20\\
      & & 5-shot & 37.22 \textpm 1.31&17.77 \textpm 0.86&17.54 \textpm 0.65&  40.51 \textpm 0.35 &\textbf{7.51} \textpm 0.15\\
      \cmidrule{2-8}
      & \multirow{4}{*}{SAR} & 0-shot & 33.50 \textpm 0.83 & 15.67 \textpm 1.01 & 15.08 \textpm 0.55 & 36.31 \textpm 0.27 & 6.66 \textpm 0.22 \\
      & & 1-shot & 36.80 \textpm 1.89&17.21 \textpm 0.96&16.54 \textpm 0.70&  37.48 \textpm 0.32 &7.02 \textpm 0.22\\
      & & 3-shot & 38.66 \textpm 1.50&17.92 \textpm 1.12&17.31 \textpm 0.85& 37.93 \textpm  0.49 &7.18 \textpm 0.20\\
      & & 5-shot & \textbf{39.35} \textpm 1.38&\textbf{18.94} \textpm 0.73&\textbf{17.95} \textpm 0.49& 38.44 \textpm 0.48 &7.32 \textpm 0.22\\    
    \bottomrule
  \end{tabular}
\end{table*}

\subsection{In-Context Learning}
\label{sec:results:icl}
Having reproduced the main results on the iKAT 2023 dataset in the previous section, this section will focus specifically on reproducing the in-context learning (ICL) experiments using the STR and SAR method, providing further insights on ~\ref{int:rq1}. Table~\ref{tab:icl_fullset_comparison} illustrates the performance of the SAR and STR methods for both BM25 and ANCE retrieval.
\begin{figure}[t!]
    \centering
    \includegraphics[width=\linewidth]{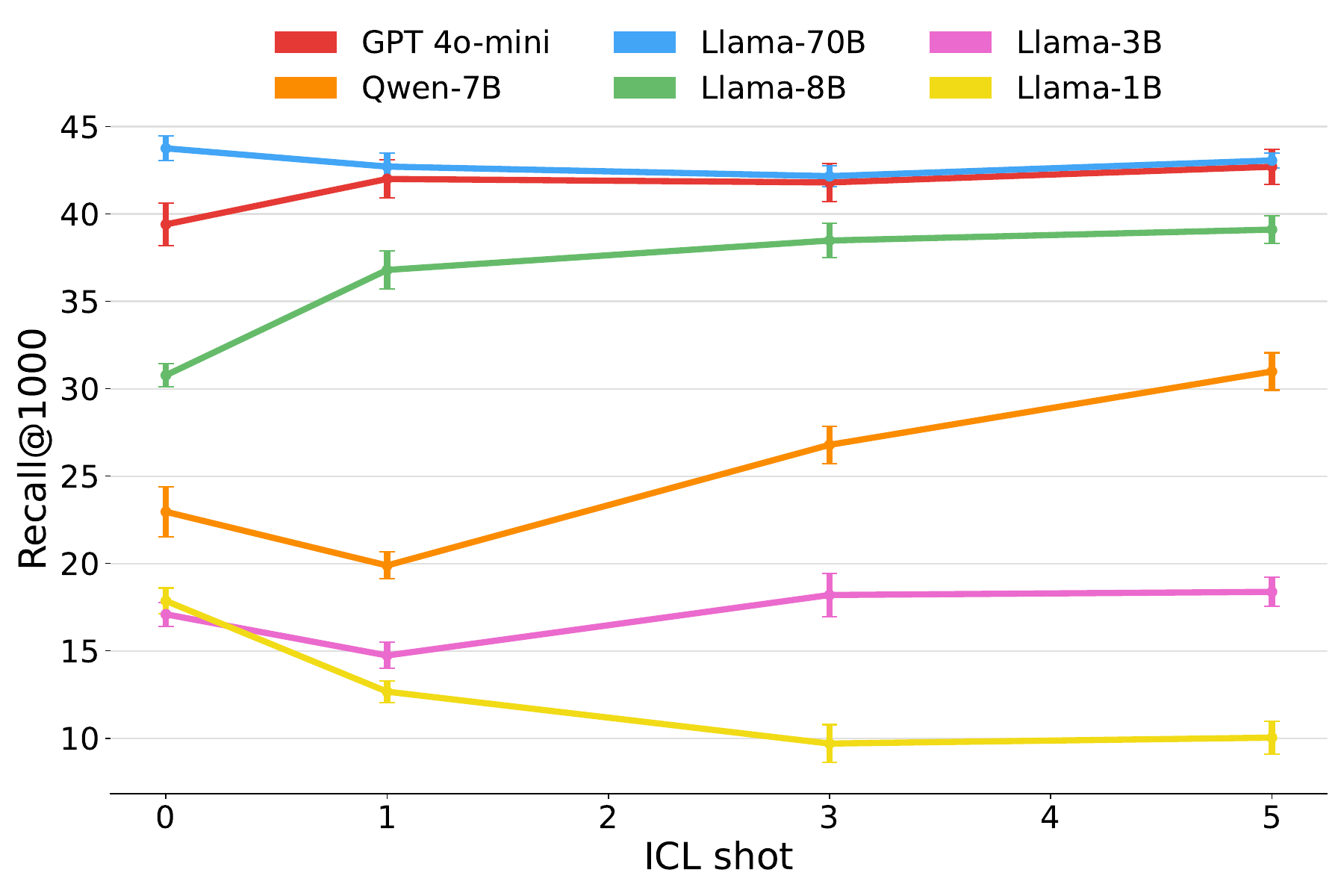}
    \caption{Open Source LLMs Performance on iKAT 2023 with varying number of in-context learning examples (Recall@1000). Retrieval with SAR, and BM25.}
    \label{fig:llama_icl_bm25}
\end{figure}
Both the None baseline and the human PTKB selection's results have also been provided again for comparison. Our results for BM25 show that neither of the STR or SAR methods is able to outperform the human PTKB selection, regardless of the number of ICL samples, with SAR significantly underperforming. For ANCE, only 5-shot SAR is able to outperform human labeling on some metrics, but not significantly. 
Interestingly, when focusing on Recall@1000, the SAR methods seem to improve in performance as we include more few-shot examples. On BM25, we see an increase from 39.64 to up to 42.70, with significant gains. Similarly, on ANCE, we observe a gain from 36.31 to 38.44 on Recall@1000, and significant gains on all metrics as well. This demonstrates that the model is able to learn from the few-shot examples, in line with findings from the original authors, on LLM guidance for properly integrating personalization.
However, when comparing the STR method, we do not see improvements with ICL. A likely reason is that the original authors included ICL examples only during the PTKB selection stage, but not in the final query reformulation prompt. In contrast, the SAR method applies ICL within a single-stage formulation, where examples are present directly in the reformulation call. We hypothesize that STR could similarly benefit from few-shot examples if they were incorporated into both stages of the pipeline.

\subsection{Open-source LLMs for Query Rewriting}
\label{sec:results:llama}
To answer~\ref{int:rq2} and assess how query rewriting performance changes when using open-source LLMs, we repeated the in-context learning experiments with the Llama and Qwen families of models. Figure~\ref{fig:llama_icl_bm25} illustrates the in-context learning results for the SAR methods using Llama 1B to 70B, Qwen 7B, and GPT-4o mini. While smaller models such as Llama-1B, Llama-3B, and Qwen-7B tend to underperform on the reformulation task, we observe that larger models perform almost on par with closed-source OpenAI's models, with the largest 70B model even outperforming GPT-4o-mini in Recall@1000. Looking back at Table~\ref{tab:fullset_comparison}, Llama-70B also outperforms the best setup of GPT-3.5, which achieves a Recall@1000 of 41.42.

Comparing the performance gains when providing some in-context examples, we see that the two smallest models (Llama-1B and Llama-3B) are unable to learn from the few-shot examples, with even performance drops. This suggests that the reformulation task is too complex and that they are unable to follow the provided instruction, together with the examples. On the other hand, we observe that both Llama-8B and Qwen-7B have gains with the few-shot examples, with Llama-8B having almost 10 points of gains. Finally, for the two best-performing models, GPT-4o-mini and Llama-70B, we do not see any gains from the in-context examples, with performances that cap around 43 points. Those results are, however, promising and show the strength of both open-source and larger LLMs for the query rewriting tasks. This also provides hints on the scaling behavior of LLMs for the query rewriting task.

\subsection{Generalization on TREC iKAT 2024}
\label{sec:results:ikat2024}

To further test the generalizability of the effectiveness of the integration of PTKB and answer ~\ref{int:rq3}, we also included TREC iKAT 2024~\cite{trecikat2024}. The results for these experiments can be found in Table~\ref{tab:full_results_2024}. The results across all metrics and methods are generally substantially higher for this dataset, indicating it is less challenging overall. Unlike the previous dataset, there are now far smaller differences between BM25 and ANCE.
Like for the 2023 dataset, human PTKB selection also significantly outperforms all methods on Recall@1000, for both BM25 and ANCE. Especially for ANCE, human PTKB selection is significantly better on most metrics compared to using all or no PTKB. These results for both iKAT 2023 and 2024 allow us to decisively state that there is a benefit to the integration of PTKB. We further compare in Section~\ref{sec:analysis:judge} the judgment pool of both datasets to understand why the results could be substantially higher compared to iKAT 2023.

\begin{table}[t!]
  \centering
  \caption{Our fivefold-averaged results for TREC iKAT 2024 using GPT-4o mini on the full test set (116 turns). STR and SAR both in zero-shot.}
  \label{tab:full_results_2024}
  \small
  \begin{tabular}{lcccc}
    \toprule
    \textbf{Method} & \textbf{MRR} & \textbf{NDCG@3} & \textbf{R@1000} & \textbf{MAP} \\
    \midrule
    \multicolumn{5}{c}{\textbf{Sparse retrieval (BM25)}} \\
    None & 68.35 \textpm 2.69 & \textbf{36.93} \textpm 1.28 & 64.33 \textpm 0.68 & 18.13 \textpm 0.49 \\
   Use All & 67.63 \textpm 2.14 & 35.63 \textpm 0.98 & 62.39 \textpm 0.72 & 17.88 \textpm 0.18 \\
   Human & \textbf{68.38} \textpm 0.26 & 36.36 \textpm 1.05 & \textbf{65.51}\textsuperscript{\textdagger} \textpm 0.62 & \textbf{18.66} \textpm 0.37 \\
  STR & 67.64 \textpm 2.75 & 36.03 \textpm 1.27 & 63.60 \textpm 0.41 & 18.56 \textpm 0.35 \\
  SAR & 60.74 \textpm 2.68 & 29.95 \textpm 1.40 & 57.21 \textpm 0.50 & 14.34 \textpm 0.41 \\
    \midrule
    \multicolumn{5}{c}{\textbf{Dense retrieval (ANCE)}} \\
    None & 66.10 \textpm 2.02 & 34.40 \textpm 1.40 & 60.13 \textpm 0.83 & 17.21 \textpm 0.16 \\
    Use All & 64.35 \textpm 3.40 & 34.80 \textpm 1.41 & 59.10 \textpm 1.21 & 17.48 \textpm 0.36 \\
    Human & \textbf{67.51} \textpm 1.42 & \textbf{37.05} \textpm 1.07 & \textbf{61.92}\textsuperscript{\textdagger} \textpm 0.86 & \textbf{18.26}\textsuperscript{\textdagger} \textpm 0.32 \\
   STR & 64.96 \textpm 2.70 & 35.44 \textpm 2.22 & 60.01 \textpm 1.05 & 17.26 \textpm 0.37 \\
    SAR & 59.60 \textpm 1.72 & 29.93 \textpm 2.16 & 55.13 \textpm 0.29 & 14.69 \textpm 0.28 \\
    \bottomrule
  \end{tabular}
\end{table}

\section{Analysis}
\label{sec:analysis}

In this section, we first analyze the agreement between the oracle and human PTKB selection, and then compare how the judgment pool can bias the performance of both BM25 and ANCE retrieval on iKAT 2023 and iKAT 2024.

\begin{table}[t]
\centering
\caption{Proportion of judged documents at different cutoffs for BM25 and ANCE with GPT-4o mini and human PTKB.}
\label{tab:judged}
\small
\begin{tabular}{lccc}
\toprule
\textbf{Retrieval} & \textbf{judged@3} & \textbf{judged@10} & \textbf{judged@100} \\
\midrule
\multicolumn{4}{c}{\textbf{iKAT 2023}} \\
\midrule
Sparse (BM25)        & \textbf{42.80} & \textbf{35.23} & \textbf{15.25} \\
Dense (ANCE)         & 37.09 & 29.02 & 11.95 \\
\midrule
\midrule
\multicolumn{4}{c}{\textbf{iKAT 2024}} \\
\midrule
Sparse (BM25)        & \textbf{65.52} & \textbf{57.41} & \textbf{28.38} \\
Dense (ANCE)         & 60.63 & 52.93 & 25.70 \\
\bottomrule
\end{tabular}
\end{table}

\subsection{Oracle Agreement with Human}
\label{sec:analysis:oracle}
\begin{figure}[t]
    \centering
    \includegraphics[width=\linewidth]{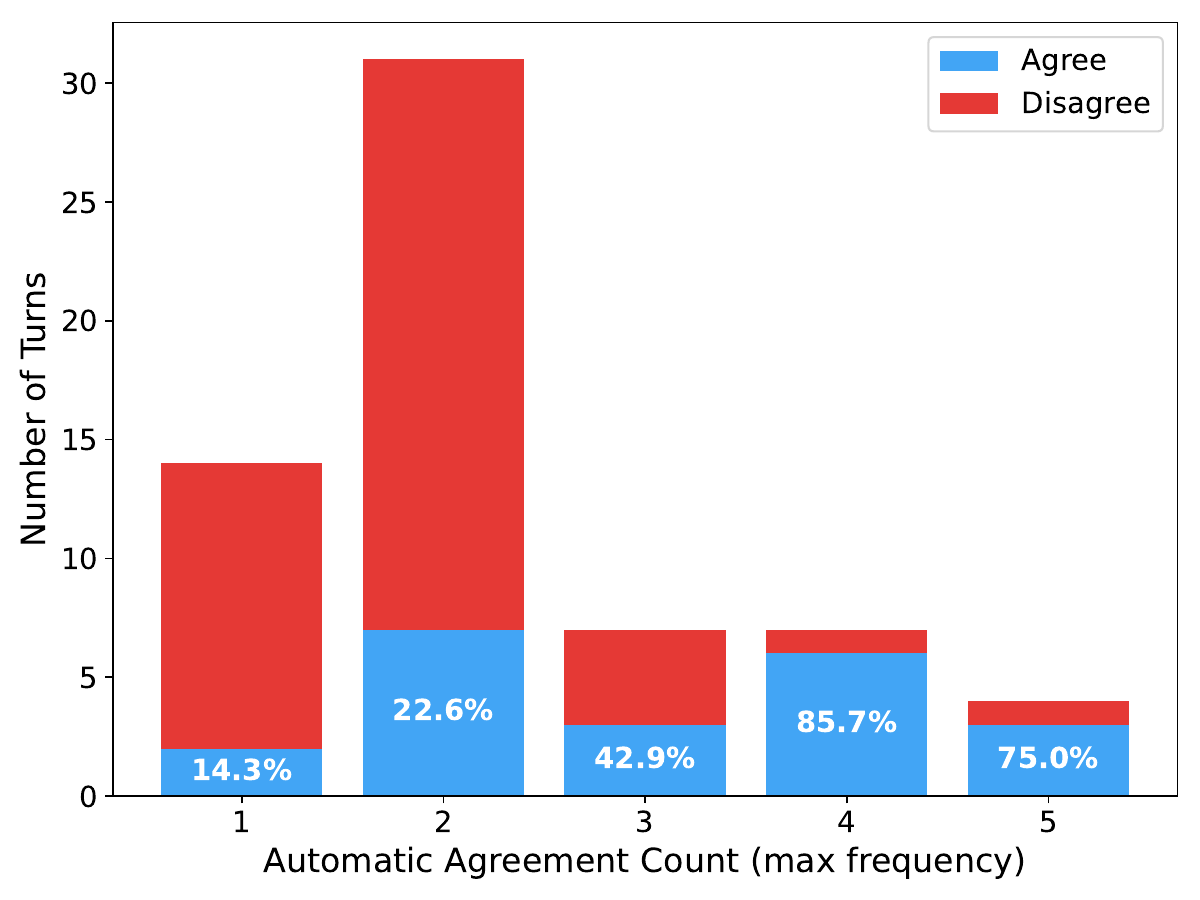}
    \caption{Agreement between human and automatic method (Recall@1000) for PTKB selection; split according to how many of the five automatic runs chose the same PTKB.}
    \Description{Agreement between human and automatic method (Recall@1000) for PTKB selection; split according to how many of the five automatic runs chose the same PTKB.}
    \label{fig:automatic_human_agreement}
\end{figure}

To quantitatively illustrate why the automatic approach is problematic, Figure~\ref{fig:automatic_human_agreement} shows how frequently the automatic method converges on a single ``best'' PTKB per turn across five runs, and whether that selection overlaps with the human-selected PTKB. If the method's performance were actually driven by effective PTKB selection, identical runs should often select the same PTKB for the same turn. However, the automatic approach seldom identifies the same PTKB consistently across the five runs, instead often selecting different PTKB elements for the same turn. 

When convergence \emph{does} occur, however, the automatic's selected PTKB typically aligns with a human-annotated PTKB element, implying that the automatic method is not inherently better than humans at identifying relevant PTKB. In turns where the method fails to converge, there may be no clear superior PTKB, and the method’s strong performance in such cases likely results from random ``lucky reformulations'' across the roughly 10 attempts, rather than any consistent advantage in PTKB selection.

\subsection{Judgment Pool of Retrieved Passages}
\label{sec:analysis:judge}

\header{BM25 vs.\ ANCE.} A key finding of both the original paper and our reproducibility study is that BM25 consistently outperforms ANCE retrieval across almost all experiments for iKAT 2023.
While this finding is likely influenced by multiple factors, the most important consideration is that the iKAT 2023 dataset is likely biased towards BM25. Specifically, only a pool of BM25 runs was used during dataset construction to select candidate passages for human annotation. Passages that were not included in this pool were automatically assigned a relevance score of zero (i.e., considered not relevant). This construction strategy inherently favors BM25, which helps explain its consistently strong performance. To assess the extent of this bias, we compare in Table~\ref{tab:judged} the number of judged passages retrieved by BM25 and ANCE at various cutoffs. The comparison confirms that BM25 retrieves a substantially higher proportion of judged documents. This suggests that some passages retrieved by ANCE may in fact be relevant, but were never presented to annotators and therefore remain unjudged. Notably, the reproduced study~\cite{mo2024pktb_cir} was the first to use dense retrieval on iKAT 2023, meaning that no dense retrieval runs were part of the original judgment pool, further complicating fair comparisons. Finally, another factor contributing to BM25's stronger performance is its access to both the LLM-generated rewrites and the LLM preliminary answers, whereas ANCE is evaluated using only the rewrite (as described in Section~\ref{sec:meth:prompt}). This difference in the input context may also explain part of the observed performance gap.

\header{TREC iKAT 2023 and iKAT 2024.}
Looking at the results on the TREC iKAT 2024 dataset, we observe that the gap between BM25 and ANCE narrows slightly, by roughly one to two points in most cutoffs, though BM25 still retains an overall lead. As shown in Table~\ref{tab:full_results_2024}, it is plausible that ANCE could outperform BM25 in some cases, but the current judgment pool may not adequately reflect its retrieval capabilities. We also find that the judgment pool for iKAT 2024 appears to be of higher quality than that of 2023. This improvement may be attributed to the widespread use of LLM-based systems in the second year of the track, which enabled the organizers to evaluate a broader range of methods.

\negskip

\section{Limitations and Future Work}
\label{sec:future}

While our reproducibility study provides initial insights into the use of LLMs for Personalized CIR, it has several limitations. One key challenge lies in the generation step: although retrieval is evaluated independently, the final generated response often serves as the primary means for personalization. For example, if a passage contains both personalized and non-personalized content, it may still be assessed as strongly relevant, even though only the personalized portion influences the generation~\cite{trecikat24conerag}. 
We also only experimented on iKAT, while real-world scenarios where user personalization is culturally nuanced or under-represented remain an open challenge~\cite{mekonnen-etal-2025-optimized}.
Besides, as this is a reproduction study, we retained the original prompts without modification. However, future work could explore prompt engineering to improve clarity and effectiveness. Since LLMs are sensitive to unclear or ungrammatical prompts \cite{cao2024worstpromptperformancelarge, promptsGPToptimising}, refining these inputs could further enhance performance and potentially reduce variance across runs.
Finally, future work could explore reranking and other paradigms for retrieval~\cite{10.5555/3600270.3601857,10.1145/3726302.3730023}.

\negskip

\section{Conclusions}
This study set out to reproduce and extend the findings of \citet{mo2024pktb_cir}, which proposed novel methods for incorporating Personal Textual Knowledge Bases (PTKBs) into personalized Conversational Information Retrieval (CIR). Notably, the original paper reported that the best performance on the iKAT 2023 dataset was achieved by omitting PTKB information during query reformulation, an unexpected result given the dataset's design. 
Our reproduction revealed substantial variability in LLM outputs, which in turn caused significant fluctuations in retrieval performance. To account for this, we repeated each experiment five times and found that many differences reported between methods were not statistically significant. Once variance was considered, the original conclusion could not be reproduced: across most metrics, all methods performed similarly within confidence intervals. However, for Recall@1000, human-annotated PTKB selections consistently outperformed other approaches.
We further showed that these findings generalize to the iKAT 2024 dataset, where human PTKB selection again yielded the best performance, significantly so on recall-based metrics for both BM25 and ANCE. This contrasts with the original study's claim that human relevance annotations were less effective than LLM-based methods.
Additionally, we evaluated open-source models of varying sizes in both zero-shot and few-shot fashion. Our results suggest that such models can effectively be used for personalized query reformulation even in more complex cases.

Overall, this work underscores the importance of rigorous experimental design when evaluating LLM-based CIR pipelines. Our results highlight the promise of PTKBs for improving user-personalized retrieval, while also emphasizing the need to account for LLM variability, particularly with commercial models that can be sensitive to prompt phrasing~\cite{errica2024did}. Robust evaluation practices, including multiple runs and testing across datasets and models, are essential to advancing reproducible and effective personalization in CIR, and its downstream applications.

\begin{acks}

This research was partly supported by the Swiss National Science Foundation (SNSF), under the project PACINO (Personality And Conversational INformatiOn Access), grant number 215742.

\end{acks}

\bibliographystyle{ACM-Reference-Format}
\balance
\bibliography{references}


\begin{thebibliography}{45}


\ifx \showCODEN    \undefined \def \showCODEN     #1{\unskip}     \fi
\ifx \showISBNx    \undefined \def \showISBNx     #1{\unskip}     \fi
\ifx \showISBNxiii \undefined \def \showISBNxiii  #1{\unskip}     \fi
\ifx \showISSN     \undefined \def \showISSN      #1{\unskip}     \fi
\ifx \showLCCN     \undefined \def \showLCCN      #1{\unskip}     \fi
\ifx \shownote     \undefined \def \shownote      #1{#1}          \fi
\ifx \showarticletitle \undefined \def \showarticletitle #1{#1}   \fi
\ifx \showURL      \undefined \def \showURL       {\relax}        \fi
\providecommand\bibfield[2]{#2}
\providecommand\bibinfo[2]{#2}
\providecommand\natexlab[1]{#1}
\providecommand\showeprint[2][]{arXiv:#2}

\bibitem[Abbasiantaeb et~al\mbox{.}(2024)]%
        {abbasiantaeb2024generatingmultiaspectqueriesconversational}
\bibfield{author}{\bibinfo{person}{Zahra Abbasiantaeb}, \bibinfo{person}{Simon Lupart}, {and} \bibinfo{person}{Mohammad Aliannejadi}.} \bibinfo{year}{2024}\natexlab{}.
\newblock \bibinfo{title}{Generating Multi-Aspect Queries for Conversational Search}.
\newblock
\showeprint[arxiv]{2403.19302}~[cs.IR]
\urldef\tempurl%
\url{https://arxiv.org/abs/2403.19302}
\showURL{%
\tempurl}


\bibitem[Abbasiantaeb et~al\mbox{.}(2025)]%
        {trecikat24conerag}
\bibfield{author}{\bibinfo{person}{Zahra Abbasiantaeb}, \bibinfo{person}{Simon Lupart}, \bibinfo{person}{Leif Azzopardi}, \bibinfo{person}{Jeffrey Dalton}, {and} \bibinfo{person}{Mohammad Aliannejadi}.} \bibinfo{year}{2025}\natexlab{}.
\newblock \showarticletitle{Conversational Gold: Evaluating Personalized Conversational Search System Using Gold Nuggets}. In \bibinfo{booktitle}{\emph{Proceedings of the 48th International ACM SIGIR Conference on Research and Development in Information Retrieval}} (Padua, Italy) \emph{(\bibinfo{series}{SIGIR '25})}. \bibinfo{publisher}{Association for Computing Machinery}, \bibinfo{address}{New York, NY, USA}, \bibinfo{pages}{3455–3465}.
\newblock
\showISBNx{9798400715921}
\href{https://doi.org/10.1145/3726302.3730316}{doi:\nolinkurl{10.1145/3726302.3730316}}


\bibitem[Aliannejadi et~al\mbox{.}(2024a)]%
        {trecikat2023}
\bibfield{author}{\bibinfo{person}{Mohammad Aliannejadi}, \bibinfo{person}{Zahra Abbasiantaeb}, \bibinfo{person}{Shubham Chatterjee}, \bibinfo{person}{Jeffrey Dalton}, {and} \bibinfo{person}{Leif Azzopardi}.} \bibinfo{year}{2024}\natexlab{a}.
\newblock \showarticletitle{TREC iKAT 2023: A Test Collection for Evaluating Conversational and Interactive Knowledge Assistants}. In \bibinfo{booktitle}{\emph{Proceedings of the 47th International ACM SIGIR Conference on Research and Development in Information Retrieval}} (Washington DC, USA) \emph{(\bibinfo{series}{SIGIR '24})}. \bibinfo{publisher}{Association for Computing Machinery}, \bibinfo{address}{New York, NY, USA}, \bibinfo{pages}{819–829}.
\newblock
\showISBNx{9798400704314}
\href{https://doi.org/10.1145/3626772.3657860}{doi:\nolinkurl{10.1145/3626772.3657860}}


\bibitem[Aliannejadi et~al\mbox{.}(2024b)]%
        {trecikat2024}
\bibfield{author}{\bibinfo{person}{Mohammad Aliannejadi}, \bibinfo{person}{Zahra Abbasiantaeb}, \bibinfo{person}{Simon Lupart}, \bibinfo{person}{Shubham Chatterjee}, \bibinfo{person}{Jeffrey Dalton}, {and} \bibinfo{person}{Leif Azzopardi}.} \bibinfo{year}{2024}\natexlab{b}.
\newblock \showarticletitle{TREC iKAT 2024: The Interactive Knowledge Assistance Track Overview}. In \bibinfo{booktitle}{\emph{The Thirty-Third Text REtrieval Conference Proceedings (TREC 2024), Gaithersburg, MD, USA, November 15-18, 2024}} \emph{(\bibinfo{series}{NIST Special Publication}, Vol.~\bibinfo{volume}{1329})}. \bibinfo{publisher}{National Institute of Standards and Technology (NIST)}.
\newblock
\urldef\tempurl%
\url{https://trec.nist.gov/pubs/trec33/papers/Overview_ikat.pdf}
\showURL{%
\tempurl}


\bibitem[Atil et~al\mbox{.}(2024)]%
        {atil2024llmstabilitydetailedanalysis}
\bibfield{author}{\bibinfo{person}{Berk Atil}, \bibinfo{person}{Alexa Chittams}, \bibinfo{person}{Liseng Fu}, \bibinfo{person}{Ferhan Ture}, \bibinfo{person}{Lixinyu Xu}, {and} \bibinfo{person}{Breck Baldwin}.} \bibinfo{year}{2024}\natexlab{}.
\newblock \bibinfo{title}{LLM Stability: A detailed analysis with some surprises}.
\newblock
\showeprint[arxiv]{2408.04667}~[cs.CL]
\urldef\tempurl%
\url{https://arxiv.org/abs/2408.04667}
\showURL{%
\tempurl}


\bibitem[Baidya et~al\mbox{.}(2025)]%
        {baidya2025behavior}
\bibfield{author}{\bibinfo{person}{Avinash Baidya}, \bibinfo{person}{Kamalika Das}, {and} \bibinfo{person}{Xiang Gao}.} \bibinfo{year}{2025}\natexlab{}.
\newblock \showarticletitle{The Behavior Gap: Evaluating Zero-shot LLM Agents in Complex Task-Oriented Dialogs}.
\newblock \bibinfo{journal}{\emph{arXiv preprint arXiv:2506.12266}} (\bibinfo{year}{2025}).
\newblock


\bibitem[Blackwell et~al\mbox{.}(2024)]%
        {LLM_evaluation}
\bibfield{author}{\bibinfo{person}{Robert~E. Blackwell}, \bibinfo{person}{Jon Barry}, {and} \bibinfo{person}{Anthony~G. Cohn}.} \bibinfo{year}{2024}\natexlab{}.
\newblock \bibinfo{title}{Towards Reproducible LLM Evaluation: Quantifying Uncertainty in LLM Benchmark Scores}.
\newblock
\showeprint[arxiv]{2410.03492}~[cs.CL]
\urldef\tempurl%
\url{https://arxiv.org/abs/2410.03492}
\showURL{%
\tempurl}


\bibitem[Cao et~al\mbox{.}(2024a)]%
        {cao2024worst}
\bibfield{author}{\bibinfo{person}{Bowen Cao}, \bibinfo{person}{Deng Cai}, \bibinfo{person}{Zhisong Zhang}, \bibinfo{person}{Yuexian Zou}, {and} \bibinfo{person}{Wai Lam}.} \bibinfo{year}{2024}\natexlab{a}.
\newblock \showarticletitle{On the worst prompt performance of large language models}.
\newblock \bibinfo{journal}{\emph{Advances in Neural Information Processing Systems}}  \bibinfo{volume}{37} (\bibinfo{year}{2024}), \bibinfo{pages}{69022--69042}.
\newblock


\bibitem[Cao et~al\mbox{.}(2024b)]%
        {cao2024worstpromptperformancelarge}
\bibfield{author}{\bibinfo{person}{Bowen Cao}, \bibinfo{person}{Deng Cai}, \bibinfo{person}{Zhisong Zhang}, \bibinfo{person}{Yuexian Zou}, {and} \bibinfo{person}{Wai Lam}.} \bibinfo{year}{2024}\natexlab{b}.
\newblock \bibinfo{title}{On the Worst Prompt Performance of Large Language Models}.
\newblock
\showeprint[arxiv]{2406.10248}~[cs.CL]
\urldef\tempurl%
\url{https://arxiv.org/abs/2406.10248}
\showURL{%
\tempurl}


\bibitem[Chann(2023)]%
        {chann2023}
\bibfield{author}{\bibinfo{person}{Sherman Chann}.} \bibinfo{year}{2023}\natexlab{}.
\newblock \bibinfo{title}{{Non-determinism in GPT-4 is caused by Sparse MoE}}.
\newblock
\urldef\tempurl%
\url{https://152334h.github.io/blog/non-determinism-in-gpt-4/}
\showURL{%
\tempurl}


\bibitem[Dalton et~al\mbox{.}(2020)]%
        {treccast19}
\bibfield{author}{\bibinfo{person}{Jeffrey Dalton}, \bibinfo{person}{Chenyan Xiong}, \bibinfo{person}{Vaibhav Kumar}, {and} \bibinfo{person}{Jamie Callan}.} \bibinfo{year}{2020}\natexlab{}.
\newblock \showarticletitle{CAsT-19: A Dataset for Conversational Information Seeking}. In \bibinfo{booktitle}{\emph{Proceedings of the 43rd International ACM SIGIR Conference on Research and Development in Information Retrieval}} (Virtual Event, China) \emph{(\bibinfo{series}{SIGIR '20})}. \bibinfo{publisher}{Association for Computing Machinery}, \bibinfo{address}{New York, NY, USA}, \bibinfo{pages}{1985–1988}.
\newblock
\showISBNx{9781450380164}
\href{https://doi.org/10.1145/3397271.3401206}{doi:\nolinkurl{10.1145/3397271.3401206}}


\bibitem[Elgohary et~al\mbox{.}(2019)]%
        {elgohary-etal-2019-unpack}
\bibfield{author}{\bibinfo{person}{Ahmed Elgohary}, \bibinfo{person}{Denis Peskov}, {and} \bibinfo{person}{Jordan Boyd-Graber}.} \bibinfo{year}{2019}\natexlab{}.
\newblock \showarticletitle{Can You Unpack That? Learning to Rewrite Questions-in-Context}. In \bibinfo{booktitle}{\emph{Proceedings of the 2019 Conference on Empirical Methods in Natural Language Processing and the 9th International Joint Conference on Natural Language Processing (EMNLP-IJCNLP)}}, \bibfield{editor}{\bibinfo{person}{Kentaro Inui}, \bibinfo{person}{Jing Jiang}, \bibinfo{person}{Vincent Ng}, {and} \bibinfo{person}{Xiaojun Wan}} (Eds.). \bibinfo{publisher}{Association for Computational Linguistics}, \bibinfo{address}{Hong Kong, China}, \bibinfo{pages}{5918--5924}.
\newblock
\href{https://doi.org/10.18653/v1/D19-1605}{doi:\nolinkurl{10.18653/v1/D19-1605}}


\bibitem[Errica et~al\mbox{.}(2024)]%
        {errica2024did}
\bibfield{author}{\bibinfo{person}{Federico Errica}, \bibinfo{person}{Giuseppe Siracusano}, \bibinfo{person}{Davide Sanvito}, {and} \bibinfo{person}{Roberto Bifulco}.} \bibinfo{year}{2024}\natexlab{}.
\newblock \showarticletitle{What Did I Do Wrong? Quantifying LLMs' Sensitivity and Consistency to Prompt Engineering}.
\newblock \bibinfo{journal}{\emph{arXiv preprint arXiv:2406.12334}} (\bibinfo{year}{2024}).
\newblock


\bibitem[Gao et~al\mbox{.}(2020)]%
        {gao2020recent}
\bibfield{author}{\bibinfo{person}{Jianfeng Gao}, \bibinfo{person}{Chenyan Xiong}, {and} \bibinfo{person}{Paul Bennett}.} \bibinfo{year}{2020}\natexlab{}.
\newblock \showarticletitle{Recent Advances in Conversational Information Retrieval}. In \bibinfo{booktitle}{\emph{Proceedings of the 43rd International ACM SIGIR Conference on Research and Development in Information Retrieval}} (Virtual Event, China) \emph{(\bibinfo{series}{SIGIR '20})}. \bibinfo{publisher}{Association for Computing Machinery}, \bibinfo{address}{New York, NY, USA}, \bibinfo{pages}{2421–2424}.
\newblock
\showISBNx{9781450380164}
\href{https://doi.org/10.1145/3397271.3401418}{doi:\nolinkurl{10.1145/3397271.3401418}}


\bibitem[Gao et~al\mbox{.}(2022)]%
        {neuralcs}
\bibfield{author}{\bibinfo{person}{Jianfeng Gao}, \bibinfo{person}{Chenyan Xiong}, \bibinfo{person}{Paul Bennett}, {and} \bibinfo{person}{Nick Craswell}.} \bibinfo{year}{2022}\natexlab{}.
\newblock \showarticletitle{Neural Approaches to Conversational Information Retrieval}.
\newblock \bibinfo{journal}{\emph{CoRR}}  \bibinfo{volume}{abs/2201.05176} (\bibinfo{year}{2022}).
\newblock
\showeprint[arXiv]{2201.05176}
\urldef\tempurl%
\url{https://arxiv.org/abs/2201.05176}
\showURL{%
\tempurl}


\bibitem[Grattafiori et~al\mbox{.}(2024)]%
        {llama3}
\bibfield{author}{\bibinfo{person}{Aaron Grattafiori}, \bibinfo{person}{Abhimanyu Dubey}, \bibinfo{person}{Abhinav Jauhri}, \bibinfo{person}{Abhinav Pandey}, \bibinfo{person}{Abhishek Kadian}, \bibinfo{person}{Ahmad Al-Dahle}, \bibinfo{person}{Aiesha Letman}, \bibinfo{person}{Akhil Mathur}, \bibinfo{person}{Alan Schelten}, \bibinfo{person}{Alex Vaughan}, \bibinfo{person}{Amy Yang}, \bibinfo{person}{Angela Fan}, \bibinfo{person}{Anirudh Goyal}, \bibinfo{person}{Anthony Hartshorn}, \bibinfo{person}{Aobo Yang}, \bibinfo{person}{Archi Mitra}, \bibinfo{person}{Archie Sravankumar}, \bibinfo{person}{Artem Korenev}, \bibinfo{person}{Arthur Hinsvark}, \bibinfo{person}{Arun Rao}, \bibinfo{person}{Aston Zhang}, \bibinfo{person}{Aurelien Rodriguez}, \bibinfo{person}{Austen Gregerson}, \bibinfo{person}{Ava Spataru}, \bibinfo{person}{Baptiste Roziere}, \bibinfo{person}{Bethany Biron}, \bibinfo{person}{Binh Tang}, \bibinfo{person}{Bobbie Chern}, \bibinfo{person}{Charlotte Caucheteux}, \bibinfo{person}{Chaya Nayak},
  \bibinfo{person}{Chloe Bi}, \bibinfo{person}{Chris Marra}, \bibinfo{person}{Chris McConnell}, {et~al\mbox{.}}} \bibinfo{year}{2024}\natexlab{}.
\newblock \bibinfo{title}{The Llama 3 Herd of Models}.
\newblock
\showeprint[arxiv]{2407.21783}~[cs.AI]
\urldef\tempurl%
\url{https://arxiv.org/abs/2407.21783}
\showURL{%
\tempurl}


\bibitem[Hambarde and Proença(2023)]%
        {hambarde2023information}
\bibfield{author}{\bibinfo{person}{Kailash~A. Hambarde} {and} \bibinfo{person}{Hugo Proença}.} \bibinfo{year}{2023}\natexlab{}.
\newblock \showarticletitle{Information Retrieval: Recent Advances and Beyond}.
\newblock \bibinfo{journal}{\emph{IEEE Access}}  \bibinfo{volume}{11} (\bibinfo{year}{2023}), \bibinfo{pages}{76581--76604}.
\newblock
\href{https://doi.org/10.1109/ACCESS.2023.3295776}{doi:\nolinkurl{10.1109/ACCESS.2023.3295776}}


\bibitem[Hui et~al\mbox{.}(2024)]%
        {hui2024qwen2}
\bibfield{author}{\bibinfo{person}{Binyuan Hui}, \bibinfo{person}{Jian Yang}, \bibinfo{person}{Zeyu Cui}, \bibinfo{person}{Jiaxi Yang}, \bibinfo{person}{Dayiheng Liu}, \bibinfo{person}{Lei Zhang}, \bibinfo{person}{Tianyu Liu}, \bibinfo{person}{Jiajun Zhang}, \bibinfo{person}{Bowen Yu}, \bibinfo{person}{Keming Lu}, {et~al\mbox{.}}} \bibinfo{year}{2024}\natexlab{}.
\newblock \showarticletitle{Qwen2. 5-coder technical report}.
\newblock \bibinfo{journal}{\emph{arXiv preprint arXiv:2409.12186}} (\bibinfo{year}{2024}).
\newblock


\bibitem[Jin et~al\mbox{.}(2023)]%
        {jin2023instructor}
\bibfield{author}{\bibinfo{person}{Zhuoran Jin}, \bibinfo{person}{Pengfei Cao}, \bibinfo{person}{Yubo Chen}, \bibinfo{person}{Kang Liu}, {and} \bibinfo{person}{Jun Zhao}.} \bibinfo{year}{2023}\natexlab{}.
\newblock \showarticletitle{{I}nstructo{R}: Instructing Unsupervised Conversational Dense Retrieval with Large Language Models}. In \bibinfo{booktitle}{\emph{Findings of the Association for Computational Linguistics: EMNLP 2023}}, \bibfield{editor}{\bibinfo{person}{Houda Bouamor}, \bibinfo{person}{Juan Pino}, {and} \bibinfo{person}{Kalika Bali}} (Eds.). \bibinfo{publisher}{Association for Computational Linguistics}, \bibinfo{address}{Singapore}, \bibinfo{pages}{6649--6675}.
\newblock
\href{https://doi.org/10.18653/v1/2023.findings-emnlp.443}{doi:\nolinkurl{10.18653/v1/2023.findings-emnlp.443}}


\bibitem[Johnson et~al\mbox{.}(2019)]%
        {faiss}
\bibfield{author}{\bibinfo{person}{Jeff Johnson}, \bibinfo{person}{Matthijs Douze}, {and} \bibinfo{person}{Herv{\'e} J{\'e}gou}.} \bibinfo{year}{2019}\natexlab{}.
\newblock \showarticletitle{Billion-scale similarity search with GPUs}.
\newblock \bibinfo{journal}{\emph{IEEE Transactions on Big Data}} \bibinfo{volume}{7}, \bibinfo{number}{3} (\bibinfo{year}{2019}), \bibinfo{pages}{535--547}.
\newblock


\bibitem[Kumar et~al\mbox{.}(2024)]%
        {kumar2024longlamp}
\bibfield{author}{\bibinfo{person}{Ishita Kumar}, \bibinfo{person}{Snigdha Viswanathan}, \bibinfo{person}{Sushrita Yerra}, \bibinfo{person}{Alireza Salemi}, \bibinfo{person}{Ryan~A Rossi}, \bibinfo{person}{Franck Dernoncourt}, \bibinfo{person}{Hanieh Deilamsalehy}, \bibinfo{person}{Xiang Chen}, \bibinfo{person}{Ruiyi Zhang}, \bibinfo{person}{Shubham Agarwal}, {et~al\mbox{.}}} \bibinfo{year}{2024}\natexlab{}.
\newblock \bibinfo{title}{Longlamp: A benchmark for personalized long-form text generation}.
\newblock


\bibitem[Lin et~al\mbox{.}(2021)]%
        {lin2021pyserini}
\bibfield{author}{\bibinfo{person}{Jimmy Lin}, \bibinfo{person}{Xueguang Ma}, \bibinfo{person}{Sheng-Chieh Lin}, \bibinfo{person}{Jheng-Hong Yang}, \bibinfo{person}{Ronak Pradeep}, {and} \bibinfo{person}{Rodrigo Nogueira}.} \bibinfo{year}{2021}\natexlab{}.
\newblock \showarticletitle{Pyserini: A Python Toolkit for Reproducible Information Retrieval Research with Sparse and Dense Representations}. In \bibinfo{booktitle}{\emph{Proceedings of the 44th International ACM SIGIR Conference on Research and Development in Information Retrieval}} (Virtual Event, Canada) \emph{(\bibinfo{series}{SIGIR '21})}. \bibinfo{publisher}{Association for Computing Machinery}, \bibinfo{address}{New York, NY, USA}, \bibinfo{pages}{2356–2362}.
\newblock
\showISBNx{9781450380379}
\href{https://doi.org/10.1145/3404835.3463238}{doi:\nolinkurl{10.1145/3404835.3463238}}


\bibitem[Lupart et~al\mbox{.}(2024)]%
        {lupart2024irlab}
\bibfield{author}{\bibinfo{person}{Simon Lupart}, \bibinfo{person}{Zahra Abbasiantaeb}, {and} \bibinfo{person}{Mohammad Aliannejadi}.} \bibinfo{year}{2024}\natexlab{}.
\newblock \showarticletitle{IRLab@ iKAT24: Learned Sparse Retrieval with Multi-aspect LLM Query Generation for Conversational Search}.
\newblock \bibinfo{journal}{\emph{arXiv preprint arXiv:2411.14739}} (\bibinfo{year}{2024}).
\newblock


\bibitem[Lupart et~al\mbox{.}(2025)]%
        {lupart2025disco}
\bibfield{author}{\bibinfo{person}{Simon Lupart}, \bibinfo{person}{Mohammad Aliannejadi}, {and} \bibinfo{person}{Evangelos Kanoulas}.} \bibinfo{year}{2025}\natexlab{}.
\newblock \showarticletitle{DiSCo: LLM Knowledge Distillation for Efficient Sparse Retrieval in Conversational Search}. In \bibinfo{booktitle}{\emph{Proceedings of the 48th International ACM SIGIR Conference on Research and Development in Information Retrieval}}. \bibinfo{pages}{9--19}.
\newblock


\bibitem[Mao et~al\mbox{.}(2024)]%
        {mao2024chatretriever}
\bibfield{author}{\bibinfo{person}{Kelong Mao}, \bibinfo{person}{Chenlong Deng}, \bibinfo{person}{Haonan Chen}, \bibinfo{person}{Fengran Mo}, \bibinfo{person}{Zheng Liu}, \bibinfo{person}{Tetsuya Sakai}, {and} \bibinfo{person}{Zhicheng Dou}.} \bibinfo{year}{2024}\natexlab{}.
\newblock \bibinfo{title}{ChatRetriever: Adapting Large Language Models for Generalized and Robust Conversational Dense Retrieval}.
\newblock
\showeprint[arxiv]{2404.13556}~[cs.IR]
\urldef\tempurl%
\url{https://arxiv.org/abs/2404.13556}
\showURL{%
\tempurl}


\bibitem[Mao et~al\mbox{.}(2023)]%
        {mao-etal-2023-large}
\bibfield{author}{\bibinfo{person}{Kelong Mao}, \bibinfo{person}{Zhicheng Dou}, \bibinfo{person}{Fengran Mo}, \bibinfo{person}{Jiewen Hou}, \bibinfo{person}{Haonan Chen}, {and} \bibinfo{person}{Hongjin Qian}.} \bibinfo{year}{2023}\natexlab{}.
\newblock \showarticletitle{Large Language Models Know Your Contextual Search Intent: A Prompting Framework for Conversational Search}. In \bibinfo{booktitle}{\emph{Findings of the Association for Computational Linguistics: EMNLP 2023}}, \bibfield{editor}{\bibinfo{person}{Houda Bouamor}, \bibinfo{person}{Juan Pino}, {and} \bibinfo{person}{Kalika Bali}} (Eds.). \bibinfo{publisher}{Association for Computational Linguistics}, \bibinfo{address}{Singapore}, \bibinfo{pages}{1211--1225}.
\newblock
\href{https://doi.org/10.18653/v1/2023.findings-emnlp.86}{doi:\nolinkurl{10.18653/v1/2023.findings-emnlp.86}}


\bibitem[Mao et~al\mbox{.}(2022)]%
        {mao2022curriculum}
\bibfield{author}{\bibinfo{person}{Kelong Mao}, \bibinfo{person}{Zhicheng Dou}, {and} \bibinfo{person}{Hongjin Qian}.} \bibinfo{year}{2022}\natexlab{}.
\newblock \showarticletitle{Curriculum Contrastive Context Denoising for Few-shot Conversational Dense Retrieval}. In \bibinfo{booktitle}{\emph{Proceedings of the 45th International ACM SIGIR Conference on Research and Development in Information Retrieval}} (Madrid, Spain) \emph{(\bibinfo{series}{SIGIR '22})}. \bibinfo{publisher}{Association for Computing Machinery}, \bibinfo{address}{New York, NY, USA}, \bibinfo{pages}{176–186}.
\newblock
\showISBNx{9781450387323}
\href{https://doi.org/10.1145/3477495.3531961}{doi:\nolinkurl{10.1145/3477495.3531961}}


\bibitem[Mekonnen et~al\mbox{.}(2025a)]%
        {mekonnen-etal-2025-optimized}
\bibfield{author}{\bibinfo{person}{Kidist~Amde Mekonnen}, \bibinfo{person}{Yosef~Worku Alemneh}, {and} \bibinfo{person}{Maarten de Rijke}.} \bibinfo{year}{2025}\natexlab{a}.
\newblock \showarticletitle{Optimized Text Embedding Models and Benchmarks for {A}mharic Passage Retrieval}. In \bibinfo{booktitle}{\emph{Findings of the Association for Computational Linguistics: ACL 2025}}, \bibfield{editor}{\bibinfo{person}{Wanxiang Che}, \bibinfo{person}{Joyce Nabende}, \bibinfo{person}{Ekaterina Shutova}, {and} \bibinfo{person}{Mohammad~Taher Pilehvar}} (Eds.). \bibinfo{publisher}{Association for Computational Linguistics}, \bibinfo{address}{Vienna, Austria}, \bibinfo{pages}{10428--10445}.
\newblock
\showISBNx{979-8-89176-256-5}
\href{https://doi.org/10.18653/v1/2025.findings-acl.543}{doi:\nolinkurl{10.18653/v1/2025.findings-acl.543}}


\bibitem[Mekonnen et~al\mbox{.}(2025b)]%
        {10.1145/3726302.3730023}
\bibfield{author}{\bibinfo{person}{Kidist~Amde Mekonnen}, \bibinfo{person}{Yubao Tang}, {and} \bibinfo{person}{Maarten de Rijke}.} \bibinfo{year}{2025}\natexlab{b}.
\newblock \showarticletitle{Lightweight and Direct Document Relevance Optimization for Generative Information Retrieval}. In \bibinfo{booktitle}{\emph{Proceedings of the 48th International ACM SIGIR Conference on Research and Development in Information Retrieval}} (Padua, Italy) \emph{(\bibinfo{series}{SIGIR '25})}. \bibinfo{publisher}{Association for Computing Machinery}, \bibinfo{address}{New York, NY, USA}, \bibinfo{pages}{1327–1338}.
\newblock
\showISBNx{9798400715921}
\href{https://doi.org/10.1145/3726302.3730023}{doi:\nolinkurl{10.1145/3726302.3730023}}


\bibitem[Mo et~al\mbox{.}(2024a)]%
        {mo-etal-2024-chiq}
\bibfield{author}{\bibinfo{person}{Fengran Mo}, \bibinfo{person}{Abbas Ghaddar}, \bibinfo{person}{Kelong Mao}, \bibinfo{person}{Mehdi Rezagholizadeh}, \bibinfo{person}{Boxing Chen}, \bibinfo{person}{Qun Liu}, {and} \bibinfo{person}{Jian-Yun Nie}.} \bibinfo{year}{2024}\natexlab{a}.
\newblock \showarticletitle{{CHIQ}: Contextual History Enhancement for Improving Query Rewriting in Conversational Search}. In \bibinfo{booktitle}{\emph{Proceedings of the 2024 Conference on Empirical Methods in Natural Language Processing}}, \bibfield{editor}{\bibinfo{person}{Yaser Al-Onaizan}, \bibinfo{person}{Mohit Bansal}, {and} \bibinfo{person}{Yun-Nung Chen}} (Eds.). \bibinfo{publisher}{Association for Computational Linguistics}, \bibinfo{address}{Miami, Florida, USA}, \bibinfo{pages}{2253--2268}.
\newblock
\href{https://doi.org/10.18653/v1/2024.emnlp-main.135}{doi:\nolinkurl{10.18653/v1/2024.emnlp-main.135}}


\bibitem[Mo et~al\mbox{.}(2024b)]%
        {mo2024pktb_cir}
\bibfield{author}{\bibinfo{person}{Fengran Mo}, \bibinfo{person}{Longxiang Zhao}, \bibinfo{person}{Kaiyu Huang}, \bibinfo{person}{Yue Dong}, \bibinfo{person}{Degen Huang}, {and} \bibinfo{person}{Jian-Yun Nie}.} \bibinfo{year}{2024}\natexlab{b}.
\newblock \showarticletitle{How to Leverage Personal Textual Knowledge for Personalized Conversational Information Retrieval}. In \bibinfo{booktitle}{\emph{Proceedings of the 33rd ACM International Conference on Information and Knowledge Management}} (Boise, ID, USA) \emph{(\bibinfo{series}{CIKM '24})}. \bibinfo{publisher}{Association for Computing Machinery}, \bibinfo{address}{New York, NY, USA}, \bibinfo{pages}{3954–3958}.
\newblock
\showISBNx{9798400704369}
\href{https://doi.org/10.1145/3627673.3679939}{doi:\nolinkurl{10.1145/3627673.3679939}}


\bibitem[Ouyang et~al\mbox{.}(2025)]%
        {ouyang2025empirical}
\bibfield{author}{\bibinfo{person}{Shuyin Ouyang}, \bibinfo{person}{Jie~M Zhang}, \bibinfo{person}{Mark Harman}, {and} \bibinfo{person}{Meng Wang}.} \bibinfo{year}{2025}\natexlab{}.
\newblock \showarticletitle{An empirical study of the non-determinism of chatgpt in code generation}.
\newblock \bibinfo{journal}{\emph{ACM Transactions on Software Engineering and Methodology}} \bibinfo{volume}{34}, \bibinfo{number}{2} (\bibinfo{year}{2025}), \bibinfo{pages}{1--28}.
\newblock


\bibitem[Overwijk et~al\mbox{.}(2022)]%
        {overwijk2022clueweb22}
\bibfield{author}{\bibinfo{person}{Arnold Overwijk}, \bibinfo{person}{Chenyan Xiong}, {and} \bibinfo{person}{Jamie Callan}.} \bibinfo{year}{2022}\natexlab{}.
\newblock \showarticletitle{ClueWeb22: 10 Billion Web Documents with Rich Information}. In \bibinfo{booktitle}{\emph{Proceedings of the 45th International ACM SIGIR Conference on Research and Development in Information Retrieval}} (Madrid, Spain) \emph{(\bibinfo{series}{SIGIR '22})}. \bibinfo{publisher}{Association for Computing Machinery}, \bibinfo{address}{New York, NY, USA}, \bibinfo{pages}{3360–3362}.
\newblock
\showISBNx{9781450387323}
\href{https://doi.org/10.1145/3477495.3536321}{doi:\nolinkurl{10.1145/3477495.3536321}}


\bibitem[Qi et~al\mbox{.}(2024)]%
        {qi2024quantifying}
\bibfield{author}{\bibinfo{person}{Zhenting Qi}, \bibinfo{person}{Hongyin Luo}, \bibinfo{person}{Xuliang Huang}, \bibinfo{person}{Zhuokai Zhao}, \bibinfo{person}{Yibo Jiang}, \bibinfo{person}{Xiangjun Fan}, \bibinfo{person}{Himabindu Lakkaraju}, {and} \bibinfo{person}{James Glass}.} \bibinfo{year}{2024}\natexlab{}.
\newblock \showarticletitle{Quantifying generalization complexity for large language models}.
\newblock \bibinfo{journal}{\emph{arXiv preprint arXiv:2410.01769}} (\bibinfo{year}{2024}).
\newblock


\bibitem[Qu et~al\mbox{.}(2020)]%
        {qu2020open}
\bibfield{author}{\bibinfo{person}{Chen Qu}, \bibinfo{person}{Liu Yang}, \bibinfo{person}{Cen Chen}, \bibinfo{person}{Minghui Qiu}, \bibinfo{person}{W.~Bruce Croft}, {and} \bibinfo{person}{Mohit Iyyer}.} \bibinfo{year}{2020}\natexlab{}.
\newblock \showarticletitle{Open-Retrieval Conversational Question Answering}. In \bibinfo{booktitle}{\emph{Proceedings of the 43rd International ACM SIGIR Conference on Research and Development in Information Retrieval}} (Virtual Event, China) \emph{(\bibinfo{series}{SIGIR '20})}. \bibinfo{publisher}{Association for Computing Machinery}, \bibinfo{address}{New York, NY, USA}, \bibinfo{pages}{539–548}.
\newblock
\showISBNx{9781450380164}
\href{https://doi.org/10.1145/3397271.3401110}{doi:\nolinkurl{10.1145/3397271.3401110}}


\bibitem[Radlinski and Craswell(2017)]%
        {frameworkcs}
\bibfield{author}{\bibinfo{person}{Filip Radlinski} {and} \bibinfo{person}{Nick Craswell}.} \bibinfo{year}{2017}\natexlab{}.
\newblock \showarticletitle{A Theoretical Framework for Conversational Search}. In \bibinfo{booktitle}{\emph{Proceedings of the 2017 Conference on Conference Human Information Interaction and Retrieval}} (Oslo, Norway) \emph{(\bibinfo{series}{CHIIR '17})}. \bibinfo{publisher}{Association for Computing Machinery}, \bibinfo{address}{New York, NY, USA}, \bibinfo{pages}{117–126}.
\newblock
\showISBNx{9781450346771}
\href{https://doi.org/10.1145/3020165.3020183}{doi:\nolinkurl{10.1145/3020165.3020183}}


\bibitem[Sabbatella et~al\mbox{.}(2024)]%
        {promptsGPToptimising}
\bibfield{author}{\bibinfo{person}{Antonio Sabbatella}, \bibinfo{person}{Andrea Ponti}, \bibinfo{person}{Ilaria Giordani}, \bibinfo{person}{Antonio Candelieri}, {and} \bibinfo{person}{Francesco Archetti}.} \bibinfo{year}{2024}\natexlab{}.
\newblock \showarticletitle{Prompt Optimization in Large Language Models}.
\newblock \bibinfo{journal}{\emph{Mathematics}} \bibinfo{volume}{12}, \bibinfo{number}{6} (\bibinfo{year}{2024}), \bibinfo{pages}{929}.
\newblock


\bibitem[Salemi et~al\mbox{.}(2025)]%
        {salemi2025expert}
\bibfield{author}{\bibinfo{person}{Alireza Salemi}, \bibinfo{person}{Julian Killingback}, {and} \bibinfo{person}{Hamed Zamani}.} \bibinfo{year}{2025}\natexlab{}.
\newblock \bibinfo{title}{ExPerT: Effective and Explainable Evaluation of Personalized Long-Form Text Generation}.
\newblock


\bibitem[Salemi et~al\mbox{.}(2023)]%
        {salemi2023lamp}
\bibfield{author}{\bibinfo{person}{Alireza Salemi}, \bibinfo{person}{Sheshera Mysore}, \bibinfo{person}{Michael Bendersky}, {and} \bibinfo{person}{Hamed Zamani}.} \bibinfo{year}{2023}\natexlab{}.
\newblock \bibinfo{title}{Lamp: When large language models meet personalization}.
\newblock


\bibitem[Tay et~al\mbox{.}(2022)]%
        {10.5555/3600270.3601857}
\bibfield{author}{\bibinfo{person}{Yi Tay}, \bibinfo{person}{Vinh~Q. Tran}, \bibinfo{person}{Mostafa Dehghani}, \bibinfo{person}{Jianmo Ni}, \bibinfo{person}{Dara Bahri}, \bibinfo{person}{Harsh Mehta}, \bibinfo{person}{Zhen Qin}, \bibinfo{person}{Kai Hui}, \bibinfo{person}{Zhe Zhao}, \bibinfo{person}{Jai Gupta}, \bibinfo{person}{Tal Schuster}, \bibinfo{person}{William~W. Cohen}, {and} \bibinfo{person}{Donald Metzler}.} \bibinfo{year}{2022}\natexlab{}.
\newblock \showarticletitle{Transformer memory as a differentiable search index}. In \bibinfo{booktitle}{\emph{Proceedings of the 36th International Conference on Neural Information Processing Systems}} (New Orleans, LA, USA) \emph{(\bibinfo{series}{NIPS '22})}. \bibinfo{publisher}{Curran Associates Inc.}, \bibinfo{address}{Red Hook, NY, USA}, Article \bibinfo{articleno}{1587}, \bibinfo{numpages}{13}~pages.
\newblock
\showISBNx{9781713871088}


\bibitem[Wang et~al\mbox{.}(2024)]%
        {temperature}
\bibfield{author}{\bibinfo{person}{Yiming Wang}, \bibinfo{person}{Ziyang Zhang}, \bibinfo{person}{Hanwei Chen}, {and} \bibinfo{person}{Huayi Shen}.} \bibinfo{year}{2024}\natexlab{}.
\newblock \bibinfo{title}{Reasoning with Large Language Models on Graph Tasks: The Influence of Temperature}.
\newblock \bibinfo{numpages}{630-634}~pages.
\newblock
\href{https://doi.org/10.1109/ICCEA62105.2024.10603677}{doi:\nolinkurl{10.1109/ICCEA62105.2024.10603677}}


\bibitem[Xiao et~al\mbox{.}(2023)]%
        {recall_neural_ir}
\bibfield{author}{\bibinfo{person}{Yan Xiao}, \bibinfo{person}{Yixing Fan}, \bibinfo{person}{Ruqing Zhang}, {and} \bibinfo{person}{Jiafeng Guo}.} \bibinfo{year}{2023}\natexlab{}.
\newblock \showarticletitle{Beyond Precision: A Study on Recall of Initial Retrieval with Neural Representations}. In \bibinfo{booktitle}{\emph{Information Retrieval: 28th China Conference, CCIR 2022, Chongqing, China, September 16–18, 2022, Revised Selected Papers}} (Chongqing, China). \bibinfo{publisher}{Springer-Verlag}, \bibinfo{address}{Berlin, Heidelberg}, \bibinfo{pages}{76–89}.
\newblock
\showISBNx{978-3-031-24754-5}
\href{https://doi.org/10.1007/978-3-031-24755-2_7}{doi:\nolinkurl{10.1007/978-3-031-24755-2_7}}


\bibitem[Xiong et~al\mbox{.}(2020)]%
        {ANCE}
\bibfield{author}{\bibinfo{person}{Lee Xiong}, \bibinfo{person}{Chenyan Xiong}, \bibinfo{person}{Ye Li}, \bibinfo{person}{Kwok-Fung Tang}, \bibinfo{person}{Jialin Liu}, \bibinfo{person}{Paul Bennett}, \bibinfo{person}{Junaid Ahmed}, {and} \bibinfo{person}{Arnold Overwijk}.} \bibinfo{year}{2020}\natexlab{}.
\newblock \bibinfo{title}{Approximate Nearest Neighbor Negative Contrastive Learning for Dense Text Retrieval}.
\newblock
\showeprint[arxiv]{2007.00808}~[cs.IR]


\bibitem[Ye et~al\mbox{.}(2023)]%
        {ye-etal-2023-enhancing}
\bibfield{author}{\bibinfo{person}{Fanghua Ye}, \bibinfo{person}{Meng Fang}, \bibinfo{person}{Shenghui Li}, {and} \bibinfo{person}{Emine Yilmaz}.} \bibinfo{year}{2023}\natexlab{}.
\newblock \showarticletitle{Enhancing Conversational Search: Large Language Model-Aided Informative Query Rewriting}. In \bibinfo{booktitle}{\emph{Findings of the Association for Computational Linguistics: EMNLP 2023}}, \bibfield{editor}{\bibinfo{person}{Houda Bouamor}, \bibinfo{person}{Juan Pino}, {and} \bibinfo{person}{Kalika Bali}} (Eds.). \bibinfo{publisher}{Association for Computational Linguistics}, \bibinfo{address}{Singapore}, \bibinfo{pages}{5985--6006}.
\newblock
\href{https://doi.org/10.18653/v1/2023.findings-emnlp.398}{doi:\nolinkurl{10.18653/v1/2023.findings-emnlp.398}}


\bibitem[Yu et~al\mbox{.}(2021)]%
        {convdr}
\bibfield{author}{\bibinfo{person}{Shi Yu}, \bibinfo{person}{Zhenghao Liu}, \bibinfo{person}{Chenyan Xiong}, \bibinfo{person}{Tao Feng}, {and} \bibinfo{person}{Zhiyuan Liu}.} \bibinfo{year}{2021}\natexlab{}.
\newblock \showarticletitle{Few-Shot Conversational Dense Retrieval}. In \bibinfo{booktitle}{\emph{Proceedings of the 44th International ACM SIGIR Conference on Research and Development in Information Retrieval}} (Virtual Event, Canada) \emph{(\bibinfo{series}{SIGIR '21})}. \bibinfo{publisher}{Association for Computing Machinery}, \bibinfo{address}{New York, NY, USA}, \bibinfo{pages}{829–838}.
\newblock
\showISBNx{9781450380379}
\href{https://doi.org/10.1145/3404835.3462856}{doi:\nolinkurl{10.1145/3404835.3462856}}


\end{thebibliography}



\end{document}